\documentclass[]{aastex63}

\graphicspath{./}

\received{August 26, 2021}
\accepted{November 4, 2021}

\shorttitle{SN~2020hvf}
\shortauthors{J. Jiang et al.}

\begin{document}

\title{Discovery of the Fastest Early Optical Emission from Overluminous SN Ia 2020hvf: A Thermonuclear Explosion within a Dense Circumstellar Environment}

\correspondingauthor{Ji-an Jiang}
\email{jian.jiang@ipmu.jp}

\author[0000-0002-9092-0593]{Ji-an Jiang}
\affil{Kavli Institute for the Physics and Mathematics of the Universe (WPI), The University of Tokyo Institutes for Advanced Study, The University of Tokyo, 5-1-5 Kashiwanoha, Kashiwa, Chiba 277-8583, Japan}

\author[0000-0003-2611-7269]{Keiichi Maeda}
\affiliation{Department of Astronomy, Kyoto University, Kitashirakawa-Oiwake-cho, Sakyo-ku, Kyoto 606-8502, Japan}

\author[0000-0002-4540-4928]{Miho Kawabata}
\affiliation{Department of Astronomy, Kyoto University, Kitashirakawa-Oiwake-cho, Sakyo-ku, Kyoto 606-8502, Japan}

\author{Mamoru Doi}
\affiliation{Institute of Astronomy, Graduate School of Science, The University of Tokyo, 2-21-1 Osawa, Mitaka, Tokyo 181-0015, Japan}
\affiliation{Research Center for the Early Universe, Graduate School of Science, The University of Tokyo, 7-3-1 Hongo, Bunkyo-ku, Tokyo 113-0033, Japan}
\affiliation{Kavli Institute for the Physics and Mathematics of the Universe (WPI), The University of Tokyo Institutes for Advanced Study, The University of Tokyo, 5-1-5 Kashiwanoha, Kashiwa, Chiba 277-8583, Japan}

\author[0000-0002-4060-5931]{Toshikazu Shigeyama}
\affiliation{Research Center for the Early Universe, Graduate School of Science, The University of Tokyo, 7-3-1 Hongo, Bunkyo-ku, Tokyo 113-0033, Japan}

\author[0000-0001-8253-6850]{Masaomi Tanaka}
\affiliation{Astronomical Institute, Tohoku University, Aoba, Sendai 980-8578, Japan}
\affiliation{Kavli Institute for the Physics and Mathematics of the Universe (WPI), The University of Tokyo Institutes for Advanced Study, The University of Tokyo, 5-1-5 Kashiwanoha, Kashiwa, Chiba 277-8583, Japan}

\author[0000-0001-8537-3153]{Nozomu Tominaga}
\affiliation{National Astronomical Observatory of Japan, National Institutes of Natural Sciences, 2-21-1 Osawa, Mitaka, Tokyo 181-8588, Japan}
\affiliation{Department of Physics, Faculty of Science and Engineering, Konan University, 8-9-1 Okamoto, Kobe, Hyogo 658-8501, Japan}
\affiliation{Kavli Institute for the Physics and Mathematics of the Universe (WPI), The University of Tokyo Institutes for Advanced Study, The University of Tokyo, 5-1-5 Kashiwanoha, Kashiwa, Chiba 277-8583, Japan}

\author[0000-0001-9553-0685]{Ken'ichi Nomoto}
\affiliation{Kavli Institute for the Physics and Mathematics of the Universe (WPI), The University of Tokyo Institutes for Advanced Study, The University of Tokyo, 5-1-5 Kashiwanoha, Kashiwa, Chiba 277-8583, Japan}

\author[0000-0001-5322-5076]{Yuu Niino}
\affiliation{Institute of Astronomy, Graduate School of Science, The University of Tokyo, 2-21-1 Osawa, Mitaka, Tokyo 181-0015, Japan}
\affiliation{Research Center for the Early Universe, Graduate School of Science, The University of Tokyo, 7-3-1 Hongo, Bunkyo-ku, Tokyo 113-0033, Japan}

\author[0000-0002-8792-2205]{Shigeyuki Sako}
\affiliation{Institute of Astronomy, Graduate School of Science, The University of Tokyo, 2-21-1 Osawa, Mitaka, Tokyo 181-0015, Japan}

\author[0000-0001-5797-6010]{Ryou Ohsawa}
\affiliation{Institute of Astronomy, Graduate School of Science, The University of Tokyo, 2-21-1 Osawa, Mitaka, Tokyo 181-0015, Japan}
\affiliation{Kiso Observatory, Institute of Astronomy, Graduate School of Science, The University of Tokyo, 10762-30 Mitake, Kiso-machi, Kiso-gun, Nagano 397-0101, Japan}

\author[0000-0001-7825-0075]{Malte Schramm}
\affiliation{Graduate School of Science and Engineering, Saitama University, Shimo-Okubo 255, Sakura-ku, Saitama-shi, Saitama 338-8570, Japan}

\author[0000-0001-9456-3709]{Masayuki Yamanaka}
\affiliation{Okayama Observatory, Kyoto University, 3037-5 Honjo, Kamogata-cho, Asakuchi, Okayama 719-0232, Japan}

\author[0000-0003-4578-2619]{Naoto Kobayashi}
\affiliation{Kiso Observatory, Institute of Astronomy, Graduate School of Science, The University of Tokyo, 10762-30 Mitake, Kiso-machi, Kiso-gun, Nagano 397-0101, Japan}
\affiliation{Institute of Astronomy, Graduate School of Science, The University of Tokyo, 2-21-1 Osawa, Mitaka, Tokyo 181-0015, Japan}
\affiliation{Laboratory of Infrared High-resolution spectroscopy (LiH), Koyama Astronomical Observatory, Kyoto Sangyo University, Motoyama, Kamigamo, Kita-ku, Kyoto 603-8555, Japan}

\author{Hidenori Takahashi}
\affiliation{Kiso Observatory, Institute of Astronomy, Graduate School of Science, The University of Tokyo, 10762-30 Mitake, Kiso-machi, Kiso-gun, Nagano 397-0101, Japan}
\affiliation{Institute of Astronomy, Graduate School of Science, The University of Tokyo, 2-21-1 Osawa, Mitaka, Tokyo 181-0015, Japan}

\author{Tatsuya Nakaoka}
\affiliation{Hiroshima Astrophysical Science Center, Hiroshima University, Higashi-Hiroshima, Hiroshima 739-8526, Japan}
\affiliation{Department of Physical Science, Hiroshima University, Kagamiyama 1-3-1, Higashi-Hiroshima 739-8526, Japan}

\author[0000-0001-6099-9539]{Koji S. Kawabata}
\affiliation{Hiroshima Astrophysical Science Center, Hiroshima University, Higashi-Hiroshima, Hiroshima 739-8526, Japan}
\affiliation{Department of Physical Science, Hiroshima University, Kagamiyama 1-3-1, Higashi-Hiroshima 739-8526, Japan}

\author[0000-0002-6480-3799]{Keisuke Isogai}
\affiliation{Okayama Observatory, Kyoto University, 3037-5 Honjo, Kamogatacho, Asakuchi, Okayama 719-0232, Japan}
\affiliation{Department of Multi-Disciplinary Sciences, Graduate School of Arts and Sciences, The University of Tokyo, 3-8-1 Komaba, Meguro, Tokyo 153-8902, Japan}

\author{Tsutomu Aoki}
\affiliation{Kiso Observatory, Institute of Astronomy, Graduate School of Science, The University of Tokyo, 10762-30 Mitake, Kiso-machi, Kiso-gun, Nagano 397-0101, Japan}
\affiliation{Institute of Astronomy, Graduate School of Science, The University of Tokyo, 2-21-1 Osawa, Mitaka, Tokyo 181-0015, Japan}

\author{Sohei Kondo}
\affiliation{Kiso Observatory, Institute of Astronomy, Graduate School of Science, The University of Tokyo, 10762-30 Mitake, Kiso-machi, Kiso-gun, Nagano 397-0101, Japan}
\affiliation{Institute of Astronomy, Graduate School of Science, The University of Tokyo, 2-21-1 Osawa, Mitaka, Tokyo 181-0015, Japan}

\author{Yuki Mori}
\affiliation{Kiso Observatory, Institute of Astronomy, Graduate School of Science, The University of Tokyo, 10762-30 Mitake, Kiso-machi, Kiso-gun, Nagano 397-0101, Japan}
\affiliation{Institute of Astronomy, Graduate School of Science, The University of Tokyo, 2-21-1 Osawa, Mitaka, Tokyo 181-0015, Japan}

\author[0000-0003-1260-9502]{Ko Arimatsu}
\affiliation{The Hakubi Center/Astronomical Observatory, Graduate School of Science, Kyoto University, Kitashirakawa-oiwake-cho, Sakyo-ku, Kyoto 606-8502, Japan}

\author[0000-0001-5903-7391]{Toshihiro Kasuga}
\affiliation{National Astronomical Observatory of Japan, National Institutes of Natural Sciences, 2-21-1 Osawa, Mitaka, Tokyo 181-8588, Japan}

\author[0000-0002-1873-3494]{Shin-ichiro Okumura}
\affiliation{Japan Spaceguard Association, Bisei Spaceguard Center, 1716-3 Okura, Bisei-cho, Ibara, Okayama 714-1411, Japan}

\author[0000-0001-7501-8983]{Seitaro Urakawa}
\affiliation{Japan Spaceguard Association, Bisei Spaceguard Center, 1716-3 Okura, Bisei-cho, Ibara, Okayama 714-1411, Japan}

\author[0000-0002-5060-3673]{Daniel E. Reichart}
\affiliation{Department of Physics and Astronomy, University of North Carolina at Chapel Hill, Campus Box 3255, Chapel Hill, NC 27599-3255, USA}

\author{Kenta Taguchi}
\affiliation{Department of Astronomy, Kyoto University, Kitashirakawa-Oiwake-cho, Sakyo-ku, Kyoto 606-8502, Japan}

\author{Noriaki Arima}
\affiliation{Institute of Astronomy, Graduate School of Science, The University of Tokyo, 2-21-1 Osawa, Mitaka, Tokyo 181-0015, Japan}
\affiliation{Department of Astronomy, Graduate School of Science, The University of Tokyo, 7-3-1 Hongo, Bunkyo-ku, Tokyo 113-0033, Japan}

\author{Jin Beniyama}
\affiliation{Institute of Astronomy, Graduate School of Science, The University of Tokyo, 2-21-1 Osawa, Mitaka, Tokyo 181-0015, Japan}
\affiliation{Department of Astronomy, Graduate School of Science, The University of Tokyo, 7-3-1 Hongo, Bunkyo-ku, Tokyo 113-0033, Japan}

\author[0000-0002-6765-8988]{Kohki Uno}
\affiliation{Department of Astronomy, Kyoto University, Kitashirakawa-Oiwake-cho, Sakyo-ku, Kyoto 606-8502, Japan}

\author{Taisei Hamada}
\affiliation{Hiroshima Astrophysical Science Center, Hiroshima University, Higashi-Hiroshima, Hiroshima 739-8526, Japan}
\affiliation{Graduate School of Advanced Science and Engineering, Hiroshima University, 1-3-1 Kagamiyama, Higashi-Hiroshima, Hiroshima 739-8526, Japan}

\vspace{10pt}

\begin{abstract}

In this Letter we report a discovery of a prominent flash of a peculiar overluminous Type Ia supernova, SN~2020hvf, in about 5 hours of the supernova explosion by the first wide-field mosaic CMOS sensor imager, the Tomo-e Gozen Camera. The fast evolution of the early flash was captured by intensive intranight observations via the Tomo-e Gozen high-cadence survey. Numerical simulations show that such a prominent and fast early emission is most likely generated from an interaction between 0.01 $M_{\odot}$ circumstellar material (CSM) extending to a distance of $\sim$$10^{13}~\text{cm}$ and supernova ejecta soon after the explosion, indicating a confined dense CSM formation at the final evolution stage of the progenitor of SN~2020hvf. Based on the CSM-ejecta interaction-induced early flash, the overluminous light curve, and the high ejecta velocity of SN~2020hvf, we suggest that the SN~2020hvf may originate from a thermonuclear explosion of a super-Chandrasekhar-mass white dwarf (``super-$M\rm_{Ch}$ WD"). Systematical investigations on explosion mechanisms and hydrodynamic simulations of the super-$M\rm_{Ch}$ WD explosion are required to further test the suggested scenario and understand the progenitor of this peculiar supernova.

\end{abstract}


\section{Introduction} \label{sec:intro}

Type Ia supernovae (SNe Ia) are widely believed to be the thermonuclear explosion of a white dwarf (WD; e.g., \citealp{hillebrandt00,hoeflich17,nomoto17,nomoto18}). Despite the great success of discovering the accelerating expansion of the universe via SN Ia observations in the 1990s \citep{perlmutter97,perlmutter99,riess98}, the origin(s) of SNe Ia are still under active debate. As a vast number of SNe Ia have been discovered in recent years via wide-field transient surveys, objects with peculiar and/or extreme properties have begun to emerge and form various subclasses of SNe Ia, such as the subluminous 91bg-like \citep{filippenko92}, 02es-like \citep{ganeshalingam12}, and 02cx-like \citep{li03} as well as the overluminous 91T/99aa-like \citep{phillips92,garavini04} and ``super-Chandrasekhar-mass" \citep{howell06,silverman11} subclasses. Among the various subclasses of SNe Ia, the so-called ``super-Chandrasekhar-mass" (``super-$M\rm_{Ch}$"\footnote{Hereafter $M\rm_{Ch}$ denotes the Chandrasekhar's limiting mass (``Chandrasekhar-mass" or ``Chandrasekhar limit") for a non-rotating carbon-oxygen WD (CO WD), i.e., $M\rm_{Ch}$ = 1.46 ($Y_e/0.5)^2~M_{\odot}$, where $Y_{e}$ is the electron mole fraction \citep{chandrasekhar39}.}) SNe Ia is the most mysterious subclass of SNe Ia because the extremely luminous (a $B$-band peak magnitude M$_{B} \lesssim -19.8$) and broad ($\Delta m_{15}(B) \lesssim 0.8$) light curve is hard to be realized via radiation from the nuclear decay of $^{56}$Ni formed by exploding a WD of $\sim$1.4 $M_{\odot}$ (called as the near-Chandrasekhar mass) in theory. Due to the low event rate of the ``super-$M\rm_{Ch}$" SNe Ia, only several ``super-$M\rm_{Ch}$" SN Ia candidates have been reported and the early-phase photometric behavior is poorly understood (\citealp{howell06,hicken07,maeda09a,yamanaka09,scalzo10,silverman11,taubenberger11,yamanaka16,chen19}; see \citealp{ashall21} for a complete sample).

In addition to explaining the high luminosity with radioactive decay of massive $^{56}$Ni via the ``super-$M\rm_{Ch}$" WD explosion (e.g., \citealp{kamiya12}), other scenarios such as the interaction with extended heavy circumstellar material (CSM; \citealp{taubenberger13}) and the asymmetric $^{56}$Ni distribution \citep{hillebrandt07} have been proposed to explain the ``super-$M\rm_{Ch}$" SN Ia with a normal WD progenitor (see \citealp{taubenberger17} and references therein). Given the difficulties of theoretically reproducing the overall properties of previously discovered ``super-$M\rm_{Ch}$" SNe Ia, it is still an open question as to whether these overluminous SNe Ia really require ``super-$M\rm_{Ch}$" WD progenitors.

Besides the ultra-high luminosity, the other remarkable feature of ``super-$M\rm_{Ch}$" SNe Ia is the persistent carbon absorption which can be detected around or even after the peak epoch (i.e., a few weeks to about one month after the SN explosion). Given that the required $^{56}$Ni masses may not necessarily be above the Chandrasekhar limit to explain the ``super-$M\rm_{Ch}$" SNe Ia, in this article, we use ``carbon-rich overluminous" SNe Ia instead of the ambiguous naming method.

In the last decade, dozens of SNe Ia were discovered within a few days of their explosions and a fraction of them show luminosity excess in the early phase. Theoretically, a prominent brightening in the first few days of the explosion can be observed in UV and optical wavelengths under specific viewing directions due to the interaction between the expanding ejecta and a nondegenerate companion star, which makes SNe Ia with additional luminosity enhancement in the early time a powerful indicator of the single-degenerate progenitor scenario \citep{kasen10,maeda14,kutsuna15}. In addition to the companion-interaction scenario, an interaction between dense CSM and SN ejecta (``CSM-ejecta interaction;" \citealp{levanon15,levanon17}), vigorous mixing of radioactive $^{56}$Ni in the outermost region of SN ejecta (``surface-$^{56}$Ni-decay;" \citealp{piro16,magee20}), and radiation from short-lived radioactive elements generated by a precursor detonation at a helium shell of the primary white dwarf (``He-shell detonation" or ``He-det"; \citealp{JJA2017,maeda18,polin19,leung20,leung21}) also predict excess emissions in the early phase. Among the previously discovered early-excess SNe Ia, two carbon-rich overluminous SNe Ia, LSQ12gpw \citep{JJA2018} and ASASSN-15pz \citep{chen19}, likely show early-excess features while inadequate early-phase photometries prevent us from identifying the physical origin of the possible early excess of the carbon-rich overluminous SNe Ia. Refer to \citet{JJA2018} for further details of well-observed early-excess SNe Ia in the past.

Here we report the earliest discovery of a carbon-rich overluminous SN Ia, SN~2020hvf, in about 5 hours from the explosion through the Tomo-e Gozen transient survey. The high-cadence observation indicates a fast prominent early-excess emission lasting for about one day after the discovery. By applying a simple ``super-$M\rm_{Ch}$" WD progenitor explosion with the CSM-interaction early-excess scenario, observational characteristics of the SN~2020hvf including the prompt early-flash and overluminous light curves can be explained reasonably well. This paper is structured as follows. An introduction of the Tomo-e Gozen transient survey and observations of SN~2020hvf are presented in Sections 2 and 3, respectively. Characteristics of the SN~2020hvf are introduced in Section 4. Modeling for both early excess and overall features of the SN~2020hvf is described in Section 5. Further discussion of the early-excess origin and the progenitor of SN~2020hvf together with our conclusions are given in Section 6. Throughout the paper we adopt the flat $\Lambda$CDM cosmology with a Hubble constant $H_{0}$ = 70 km s$^{-1}$ Mpc$^{-1}$, $\Omega_{M}=0.3$.

\section{The Tomo-e Gozen Transient Survey}

The Tomo-e Gozen camera (``Tomo-e"), is a wide-field mosaic CMOS sensor imager mounted on the 1.05m Schmidt Telescope at the Kiso Observatory, the University of Tokyo, Japan. Tomo-e is developed for time-domain astronomy, equipped with 84 blue-sensitive CMOS image sensors of $2000\times1128$ pixels in size developed by Canon Inc\footnote[1]{https://global.canon/en/technology/tomoegozen2019.html}. The sensitive wavelength range of the CMOS sensor is 3800--7000 \text{\AA} with a peak at $\sim$5000 \text{\AA} \citep{sako16,sako18,kojima18}. The field of view for each CMOS sensor is $39'.7 \times 22'.4$ with an angular pixel scale of $1''.19$, corresponding to a total field of view of $\sim$20 deg$^2$. Tomo-e offers a sequential shooting mode at a maximum frame rate of 2 fps (frames per second) with a rolling shutter in whole field. The readout overhead time between exposures is in 0.1 milliseconds. The readout of the image sensor is synchronized with the GPS time and the time stamp of Tomo-e is as accurate as 0.2 milliseconds \citep{sako18}.

The Tomo-e Gozen transient survey was officially launched on 2019 October 1. Two different survey modes, a 2$\pi/$3 survey ($\sim$7000 deg$^2$) and a high-cadence survey to repeatedly observe a few thousand deg$^2$ sky are regularly conducted every night (N. Tominaga et al., in preparation). Observations are carried out without a filter (the bandpass of which is comparable to Pan-STARRS $g+r$, denoted by ``clear" in related figures). Photometries of typical extragalactic transients (e.g., supernovae, active galactic nuclei, tidal disruption events) are performed by using 6 s exposures coadded by 12 consecutive frames (2 fps), reaching to a 5$\sigma$ limiting magnitude of $\sim$18 mag in photometric dark nights. In addition, each frame (0.5 s single exposure) will be used to search fast-moving objects such as near-Earth objects, meteors \citep{ohsawa19,ohsawa20}, and Kuiper Belt objects \citep{arimatsu19} as well as unknown optical transients with variable timescales of seconds. The survey takes about 30 terabytes of data every night, $\sim$1.5 times larger than that taken by the Vera C. Rubin Observatory/LSST \citep{ivezic19}.

\section{Discovery and Follow-up Observations of SN~2020hvf}

SN~2020hvf was discovered by the Asteroid Terrestrial-impact Last Alert System (ATLAS; \citealp{tonry18}) as a supernova candidate on UT 2020 April 21.38 \citep{smith20}, which was independently discovered by Tomo-e (internally designated as Tomo-e202004aaelb) on UT 2020 April 20.5019 (MJD 58959.5019, $\sim$19.20 days before the polynomial-fitted $B$-band peak epoch), about one day earlier than the ATLAS discovery. The SN~2020hvf is located at $\alpha$(J2000) = $11^{h}21^{m}26^{s}.45$ and $\delta$(J2000) = $+03^\circ00'52''.85$, about $22''.03$ (to the northeast) from the host-galaxy center, with a nonfilter photometry of $\sim$16.5 mag upon discovery. The Tomo-e photometry of SN~2020hvf shows an abnormal variance in the first two days after the discovery, indicating a very fast early-excess emission of the transient.

The Tomo-e observation of the SN~2020hvf was continued for 10 days and then terminated due to technical maintenance of the camera. We carried out ground-based follow-up observations starting from MJD 58963.4, about 3.9 days after the Tomo-e discovery with the Seimei 3.8 m telescope of Kyoto University \citep{kurita10,kurita20} and Kanata 1.5 m telescope of Hiroshima University. $BVRIJHK_{s}$-band imaging observations were performed with the Hiroshima One-shot Wide-field Polarimeter (HOWPol; \citealp{kawabata08}) and Hiroshima Optical and Near-InfraRed camera (HONIR; \citealp{akitaya14}) installed on the Nasmyth and the Cassegrain foci of the Kanata telescope, respectively. In addition, multiband follow-up observations were performed using several small robotic telescopes from 10 days after the Tomo-e discovery. We used the 35 cm telescope DSO-14 located at the Dark Sky Observatory and the 60 cm Rapid Response Robotic Telescope (RRRT) located at the Fan Mountain Observatory. Both telescopes are operated as part of Skynet (see \citealp{martin19} for details about the network). In addition we used two 70 cm telescopes located at the Sierra Remote Observatory and Spring Brook Observatory which are operated as part of the Thai Robotic Telescope network.

\begin{figure*}
\plotone{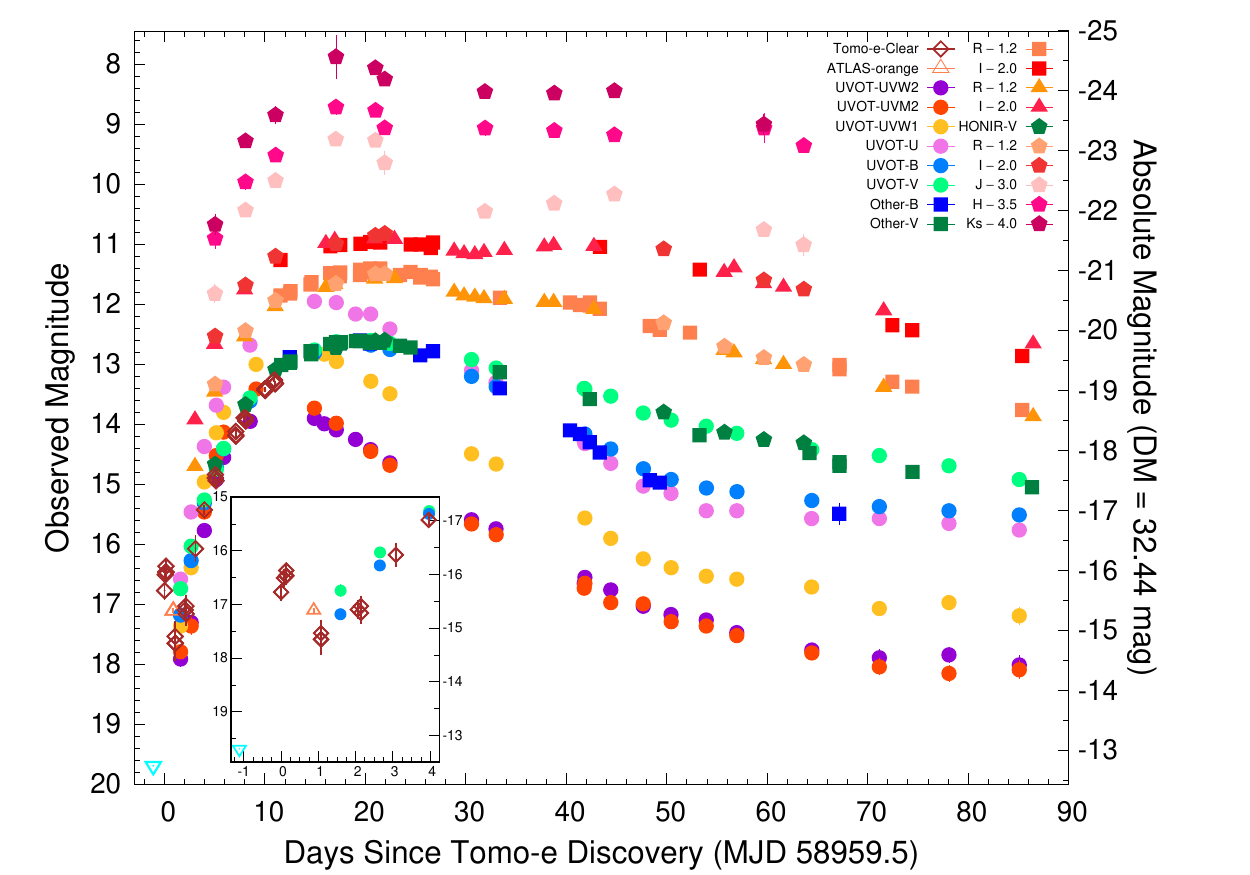}
\caption{The multiband light curves of SN~2020hvf. An inverted open triangle is the ATLAS nondetection (cyan band) on MJD 58958.378. Days since Tomo-e first detection are in the rest frame. The sub-panel zooms in on the light curves from about $-1$ to 4 days after the Tomo-e discovery. Magnitudes of Tomo-e nonfilter (open diamonds) and ATLAS-orange (an open triangle; \citealp{tonry20}) are in the AB system, while other magnitudes are in the Vega system. Circles, triangles, pentagons, and squares correspond to photometries from Swift/UVOT, Kanata/HOWPol, Kanata/HONIR, and small robotic telescopes, respectively.
\label{fig:sn2020hvfImg}}
\end{figure*}

The imaging data of Tomo-e and follow-up observations were reduced in a standard manner for the photometry. Photometric calibrations of Tomo-e data were done against the Pan-STARRS1 (PS1) $3\pi$ catalog \citep{tonry12,magnier13} by taking into account the response functions of the CMOS sensor and those of the PS1 $g$-, $r$-, and $i$-band data via the Tomo-e transient pipeline. The zero point was measured in a reference image made by Tomo-e observations at the same field before 2020 April. Aperture photometry was then performed after image subtraction by matching the scale to the reference image.

We adopted the point-spread function (PSF) fitting photometry method using DAOPHOT package in IRAF\footnote[2]{IRAF is distributed by the National Optical Astronomy Observatory, which is operated by the Association of Universities for Research in Astronomy (AURA) under a cooperative agreement with the National Science Foundation.} for the HOWPol and HONIR photometries. For the magnitude calibration of HOWPol data, we adopted relative photometry using the comparison stars. The magnitudes of the comparison stars in the $BVRI$ bands were calibrated with stars in the same field observed on a photometric night. First-order color-term correction was applied in the photometry. Sky background subtraction was applied for the HONIR NIR data by using a template sky image obtained by the dithering observation. Then we performed PSF photometry and calibrated the magnitude using the comparison stars in the 2MASS catalog \citep{persson98}.

For optical follow-up observations with small robotic telescopes, we obtained PSF photometry after decomposing the host galaxy NGC~3643 from SN~2020hvf using GALFIT \citep{galfit10}. First we constructed an empirical PSF model from field stars and then fitted a combined Sersic model (for NGC~3643) and PSF model (for SN~2020hvf). We keep the Sersic parameters fixed for each filter and telescope. The bright star $45''$ southeast from SN~2020hvf is used as a control star for the photometry.

The Ultraviolet Optical Telescope (UVOT; \citealp{roming05}) installed on the Neil Gehrels Swift Observatory began observing SN~2020hvf on MJD 58961.10, about 1.60 days after the Tomo-e discovery. The UVOT observation lasted for about three months (from $\sim$$-18$ days to 60 days after the $B$-band peak). The SN flux was measured via aperture photometry on UVOT images via the usual procedures in HEASoft, including corrections for coincidence loss and aperture loss. The image counts were converted to physical fluxes using the latest calibration \citep{breeveld11} and finally transferred to the UVOT Vega photometry system. We did not perform the image subtraction as there are no pre-SN UVOT images at the SN location in the Swift archive. Visual inspection of the UVOT images suggests a negligible host-galaxy contamination for the UVOT flux measurements. 

All magnitudes except for the Tomo-e data are given in the Vega system. The ground-based optical/NIR photometries and the Swift/UVOT photometries are given in Tables \ref{tab:Ground Photometry Data} and \ref{tab:Space Photometry Data} in the Appendix, respectively.

SN~2020hvf was spectroscopically identified as an SN Ia by \citet{burke20}, who took a spectrum of SN~2020hvf by FLOYDS-N installed on the Las Cumbres Observatory 2m telescopes on Haleakala (FTN) on MJD 58961.25, about one day after the ATLAS discovery. In our follow-up observations, a series of spectra of SN~2020hvf have been taken by the Kyoto Okayama Optical Low-dispersion Spectrograph with an integral field unit (KOOLS-IFU; \citealp{yoshida05,matsubayashi19}) mounted on the newly built Seimei 3.8 m telescope \citep{kurita10} and HOWPol from $\sim$$-16$ days to 208 days after the $B$-band peak. KOOLS-IFU is equipped with four grisms, among which we used the VPH-blue. The wavelength coverage is 4000--8900 \text{\AA} and the spectral resolution $R \sim500$. Data reduction was performed using a reduction software specifically developed for KOOLS-IFU and the Hydra package in IRAF \citep{barden94,barden95}. 

\section{Characteristics of SN~2020hvf}

\begin{figure*}
\gridline{\fig{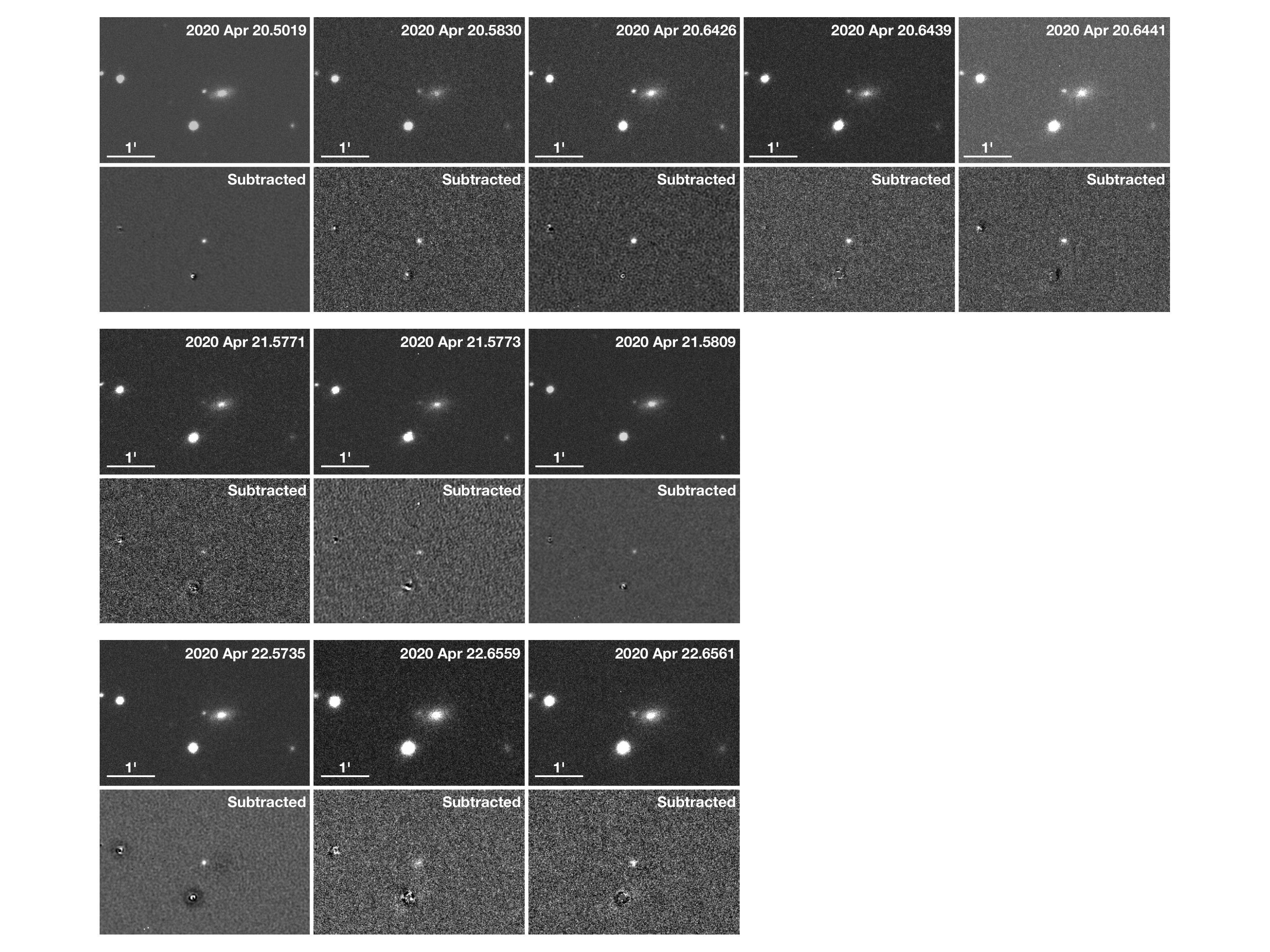}{0.66\textwidth}{}}
\vspace{-15pt}
\caption{Tomo-e images of SN~2020hvf in the first three nights. Thumbnails in odd rows are images taken in the first, second, and third nights, respectively. Corresponding reference-subtracted images are shown in even rows.
\label{fig:20hvf_thumbnails}}
\end{figure*}

Figure \ref{fig:sn2020hvfImg} presents UVOIR multiband light curves of SN~2020hvf. The first-night Tomo-e observation indicates that the brightness of the SN increased by $\sim$$50_{-39}^{+39}\%$ in 3.4 hours. However, the rising behavior did not continue and the brightness of the SN dramatically dropped by $\gtrsim$$70\%$ in one day and then got brighter again from the third day. Images of SN~2020hvf in the first three nights are presented in Figure \ref{fig:20hvf_thumbnails}. As can be seen in the figure, the discovered intraday variability is independent of observing conditions and all detections are higher than the 3$\sigma$ detection limit. Specifically, a distinct brightness variation in the first two days can be distinguished by either comparing with nearby stars from the original images or the scaled subtracted images. In addition, we did an independent test by using PSF photometry on the original images (i.e., without image subtraction) taken in the first two nights and the result shows a good consistency with our photometric result from subtracted images, further demonstrating the very fast early excess of SN~2020hvf. A deep nondetection by ATLAS one day before the Tomo-e discovery also confirms the extremely short duration of the early flash, and the onset time of the early flash given by our best-fit model is about 0.2 day before the Tomo-e discovery (Section 5). The prominent pulse-like feature lasted for about 1 day after the discovery of SN~2020hvf. Such a short-lived but bright early optical excess has never been discovered in SNe Ia so far, suggesting a different origin of the early excess compared to previously discovered early-excess SNe Ia. After the early-excess phase, the light curve of SN~2020hvf keeps rising monotonically for about 18 days and finally reaches to a very high peak luminosity.

A polynomial fitting of the $B$-band light curve around the maximum gives a $B$-band peak magnitude of $\sim$12.67 at MJD 58979.3 and ${\Delta}m_{15}(B) \sim 0.78$ for SN~2020hvf. The foreground Galactic extinction toward SN~2020hvf is $E(B-V)\rm_{MW} = 0.0356$ mag \citep{schlafly11}. A Galactic extinction-corrected $B-V$ color of $\sim$$-0.03$ at the $B$-band peak is similar to other carbon-rich overluminous SNe Ia \citep{scalzo10,silverman11,taubenberger11} and normal SNe Ia at the same phase. Nondetections of the Na$\rm_{I}$D doublet in our spectra and a far location of the SN relative to the host center suggest a negligible local extinction of the SN~2020hvf.

\begin{figure*}
\plotone{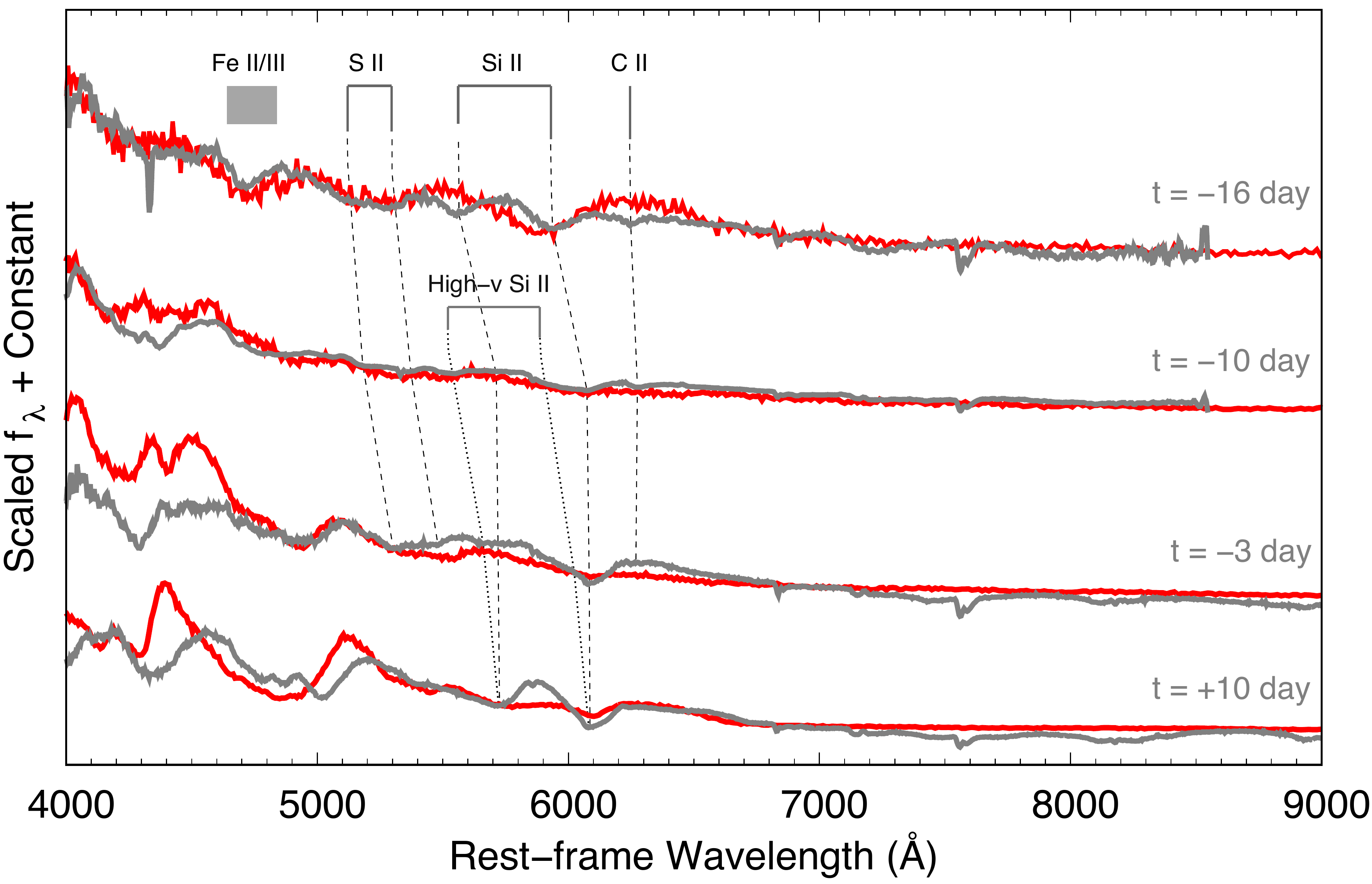}
\vspace{0pt}
\caption{Spectra of SN~2020hvf at $-16$, $-10$, $-3$, and $+10$ days relative to the $B$-band peak epoch. Galactic extinction has been corrected. Main absorption features are denoted on the top of the figure. The overall spectra show high-velocity evolution compared to previously discovered carbon-rich overluminous SNe Ia. A high-velocity component of Si $\rm _{II}$ lines showed up at $t\sim-10$ day. Spectra in red are synthesized based on our fiducial model (dashed lines in Figure \ref{fig:Overall_LC_Fitting}).
\label{fig:sn2020hvf_spec_evo}}
\end{figure*}

The redshift of the host galaxy is 0.00581 $\pm$ 0.00001 \citep{bolton12}. Using the NASA/IPAC Extragalactic Database (NED),\footnote[3]{http://ned.ipac.caltech.edu} we determine a distance modulus (DM) of $32.45 \pm 0.15$ mag to the source with corrections for peculiar velocities due to the Virgo Supercluster, Great Attractor, and Shapley Supercluster. The $B$-band peak absolute magnitude of $-19.92 \pm 0.20$ mag and a ${\Delta}m_{15}(B) \sim 0.78$ mag indicate an ultra-high luminosity of the SN~2020hvf. Even though the distance of the host galaxy NGC~3643 has an uncertainty due to its nearby location, the small ${\Delta}m_{15}(B)$ and long rise time of the SN~2020hvf indicate a significantly larger amount of $^{56}$Ni than that of normal SNe Ia and are consistent with the peak absolute magnitude derived based on the redshift-dependent DM shown here \footnote[4]{There remains a possibility that the SN~2020hvf may not obey the Phillips relation \citep{phillips92}, i.e., a very slow-evolving light curve but with normal peak brightness, while SNe Ia with these kinds of light-curve features have not been clearly confirmed.}. Given the clear intermediate-mass-element (IME) absorptions from the early time and the long-lasting C $\rm _{II}$ feature, the SN~2020hvf show a high spectral similarity to carbon-rich overluminous SNe Ia rather than 91T/99aa-like overluminous SNe Ia. We thus classify the SN~2020hvf as a carbon-rich overluminous SN Ia in this paper. We note that another estimate on the distance to the host galaxy, NGC~3643, could be obtained through a possible galaxy group membership. NGC~3643 may be associated with NGC~3640 \citep{madore04}, for which the latest measurement suggests DM $=32.15 \pm 0.14$ \citep{tully13}. This is somewhat smaller than the fiducial value we adopt, but adopting this value would still place SN~2020hvf in the overluminous category. Given that the group membership has not been established, we adopt the Hubble-flow distance (corrected for the peculiar velocity) as our fiducial value.

As shown in Figure \ref{fig:sn2020hvf_spec_evo}, the overall spectral features such as the C $\rm _{II}$ absorption lasting for more than two weeks after the discovery are in line with other carbon-rich overluminous SNe Ia (\citealp{hicken07,scalzo10,silverman11,taubenberger11}). Although the large ejecta velocity scatter has been found in this peculiar SN Ia branch, SN~2020hvf shows a significantly faster velocity evolution in the pre-max phase, which has been found only in one carbon-rich overluminous SN Ia, ASASSN-15pz \citep{chen19} in the past. Moreover, an even higher velocity component of Si $\rm _{II}$ lines was observed at $t\sim-10$ day, which was merged with the main absorption around the maximum. A detailed analysis of the full spectral data set will be given in M. Kawabata et al. (in preparation). In the following, we introduce modeling of both the prominent early optical flash and the overall observational characteristics of the SN~2020hvf.

\section{SN~2020hvf Modeling}

\subsection{Early-excess Modeling}

\begin{figure*}
\gridline{\fig{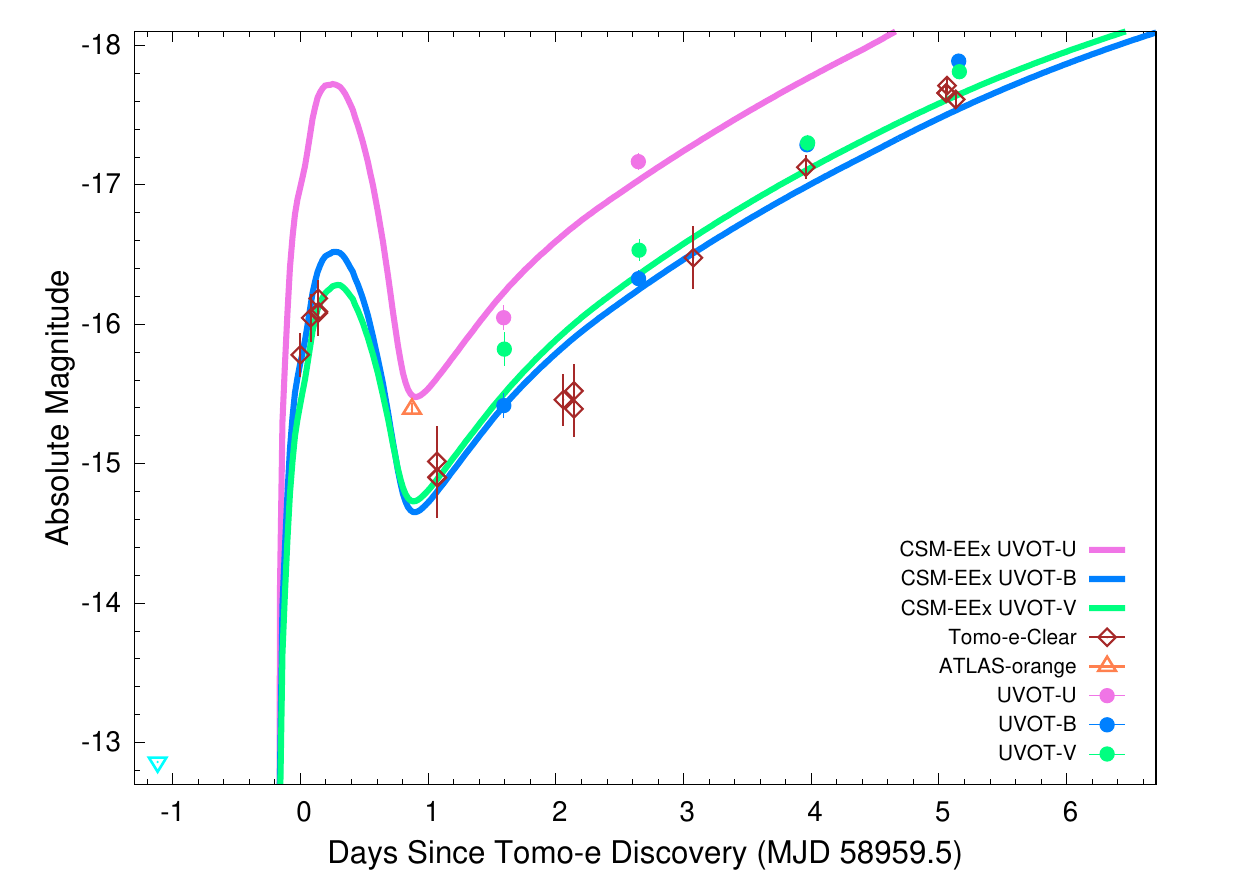}{0.5\textwidth}{(a) CSM-ejecta Interaction}
          \fig{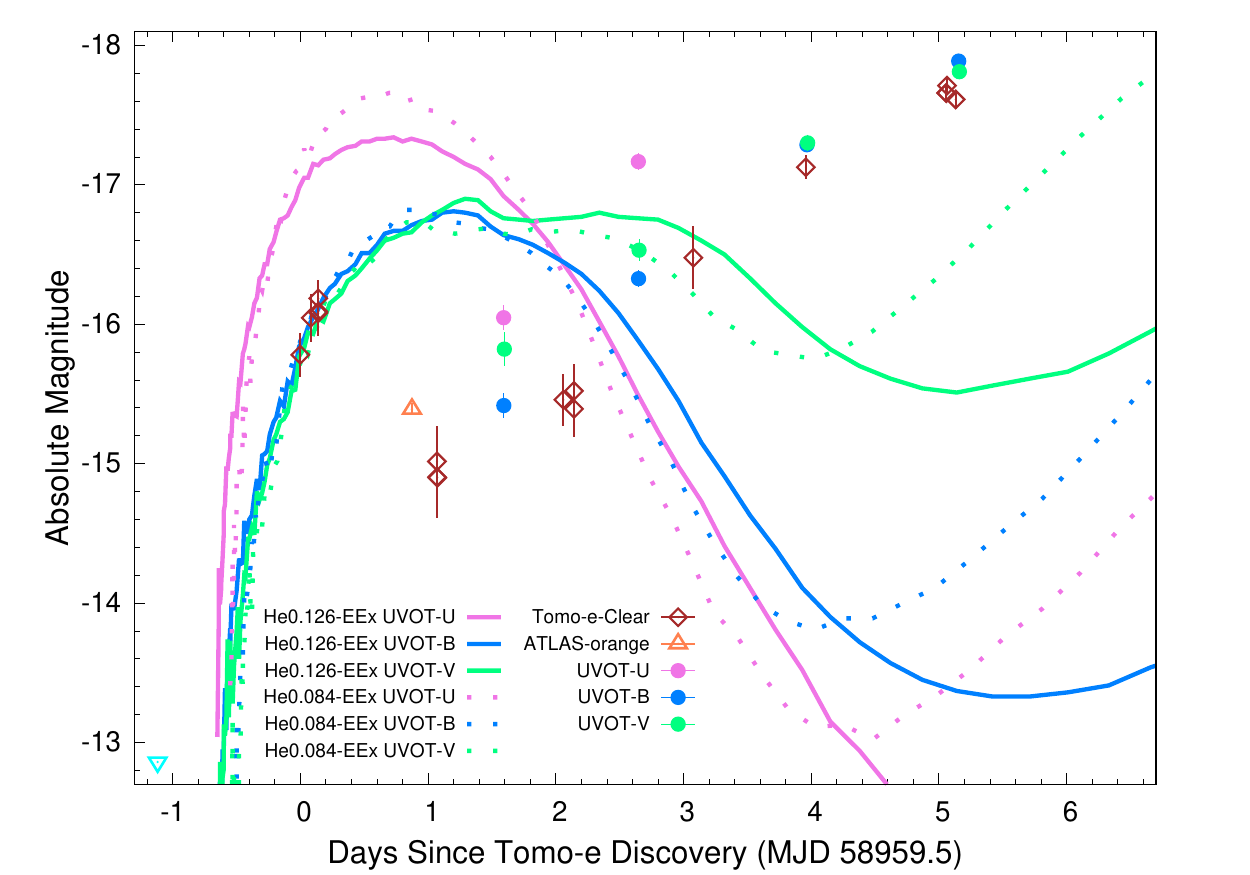}{0.5\textwidth}{(b) He-det}
          }
\gridline{\fig{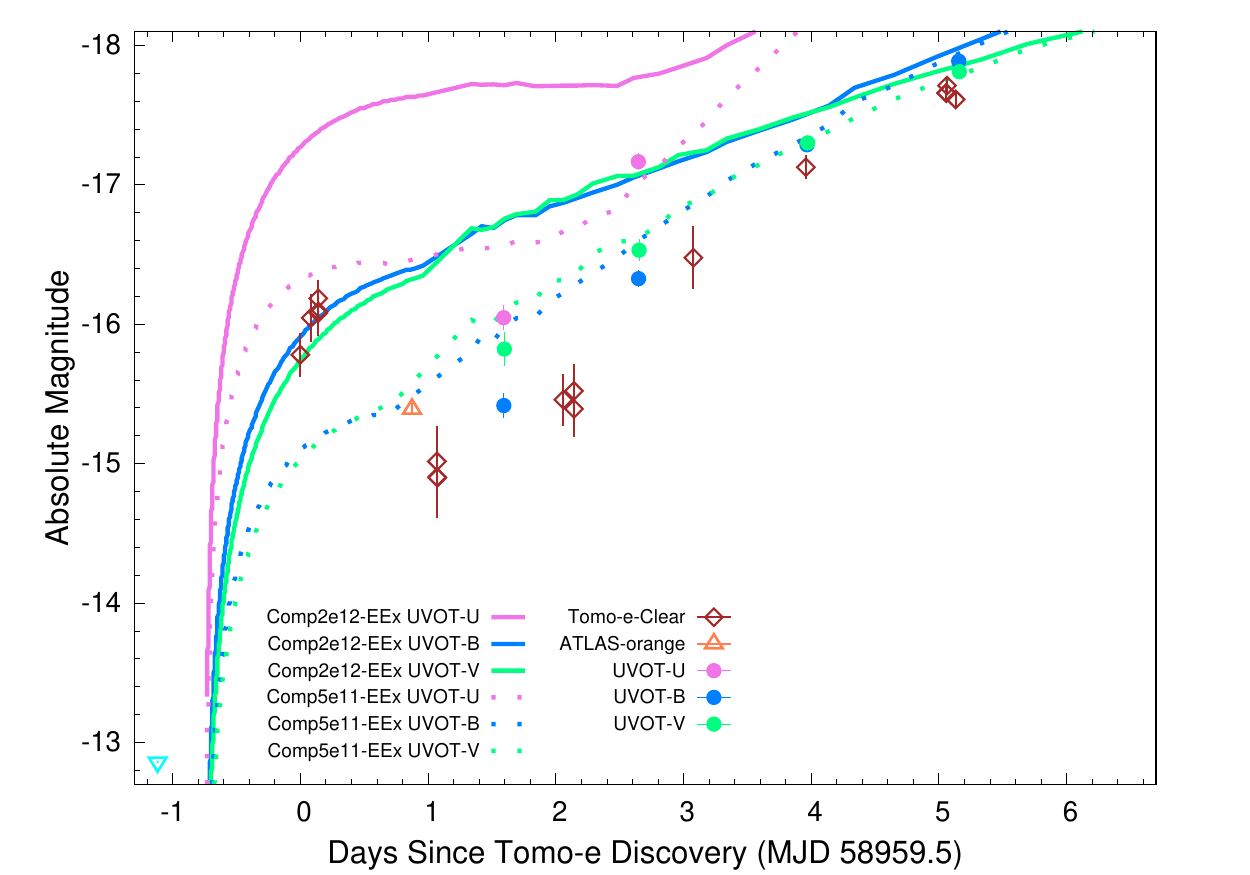}{0.5\textwidth}{(c) Companion-ejecta Interaction}
          \fig{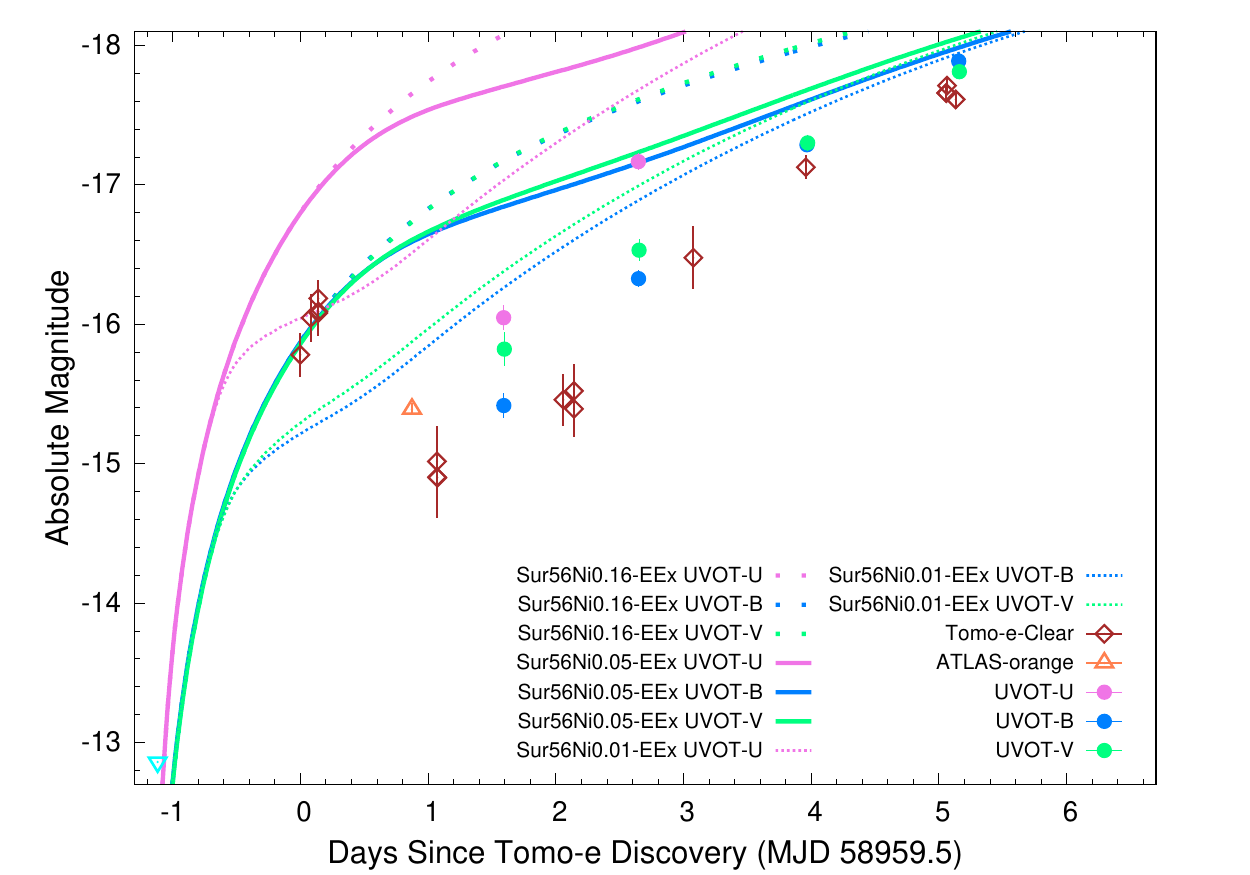}{0.5\textwidth}{(d) Surface-$^{56}$Ni Decay}
          }
\caption{Comparisons between early-phase observations of SN~2020hvf and model light curves from different early-excess scenarios. (a) Synthesized light curves of 1D CSM-ejecta interaction assuming CSM mass of $0.01~M_{\odot}$ with a characteristic (outer edge) radius of $1.0\times10^{13}$ cm (``CSM-EEx"). (b) 1D He-det early-excess simulations assuming a $0.126~M_{\odot}$ (``He0.126-EEx;" solid lines) and a $0.084~M_{\odot}$ (``He0.084-EEx;" dotted lines) He shell, respectively. (c) 1D companion-ejecta interaction simulations assuming separations between a companion star and the SN ejecta center of $2.0\times10^{12}$ cm (``Comp2e12-EEx;" solid lines) and $5.0\times10^{11}$ cm (``Comp5e11-EEx;" dotted lines), respectively. (d) Synthesized light curves of a 1D surface-$^{56}$Ni-decay scenario with $0.16~M_{\odot}$ (``Sur56Ni0.16-EEx;" sparsely dotted lines), $0.05~M_{\odot}$ (``Sur56Ni0.05-EEx;" solid lines), and $0.01~M_{\odot}$ (``Sur56Ni0.01-EEx;" dotted lines) $^{56}$Ni distributed from surface to specific velocity layers of the ejecta, respectively. Explosion time of the model in each panel is shifted ((a) $-0.2$, (b) $-0.7$, (c) $-0.75$, and (d) $-1.2$ days) to fit the first-night Tomo-e observation. \label{fig:EarlyLightCurveFittings}}
\end{figure*}

The early-excess SN Ia was originally proposed as a powerful indicator of the single-degenerate progenitor system because the interaction between a nondegenerate companion star (e.g., red giant, main-sequence star) and SN ejecta may cause prominent brightness excess in the first few days of the explosion. Recently, thanks to the progress of numerical simulations for early-phase light curves and remarkable discoveries of the early-excess SNe Ia, another three scenarios, i.e., the He-det, the CSM-ejecta interaction, and the surface-$^{56}$Ni-decay scenarios, have also been proposed to explain previous early-excess SNe Ia. In order to understand the prompt pulse-like early excess of SN~2020hvf, we systematically investigate the four early-excess scenarios. A few representative models for each scenario are shown in a panel of Figure \ref{fig:EarlyLightCurveFittings}.

We firstly introduce a ``confined" CSM and compute the interaction between the SN ejecta and the CSM. The prescriptions here are similar to \citet{piro16}. The CSM composition is assumed to be dominated by C + O with solar metals. Adopting a power-law density distribution as a function of radius with an index of $-3$, the properties of the CSM are specified by the CSM mass and outer radius. The ejecta properties are fixed by modeling the post-flash behaviors, and here we adopt 2.1 $M_\odot$ for the ejecta mass (i.e., ``super-$M\rm_{Ch}$" WD) and $1.4 \times 10^{51}$ erg for the kinetic energy (see Section 5.2). Now that the ejecta and CSM properties are specified, the system is evolved with the radiation-hydrodynamic mode of SNEC (the SuperNova Explosion Code, \citealp{morozova15}), which provides multiband light curves powered by the CSM interaction. As shown in panel (a) of Figure \ref{fig:EarlyLightCurveFittings}, a model assuming a CSM mass of $0.01~M_{\odot}$ with a characteristic (outer edge) radius of $1.0\times10^{13}$ cm can nicely fit the prominent early flash of SN~2020hvf. The CSM outer radius mostly affects the luminosity and the color, while the CSM mass mainly affects the duration \citep{piro16,maeda18}. As such, observational data can be used to strongly constrain the nature of the CSM in this scenario.

In terms of other early-excess scenarios, we find that the other three scenarios would not reproduce the observed combination of the high luminosity and the short duration of the early excess found for SN~2020hvf. In the following, we briefly introduce how the model parameters in each scenario are set to produce the early excess as bright as observed. These models are shown in Figure \ref{fig:EarlyLightCurveFittings}, demonstrating that either the timescale or characteristic evolution is not compatible with the observation of SN~2020hvf. Difficulties of interpreting the early excess of SN~2020hvf by the other three scenarios are further discussed by taking into account the discrepancies of the overall features/conditions expected by early-excess-related explosion mechanisms (e.g., He-det) and progenitor systems (e.g., companion-ejecta interaction) in Section 6.1.

For the He-detonation scenario, we examine a grid of 1D models computed previously \citep{maeda18} that covers the detonated He-shell mass of 0.003--0.13 $M_\odot$. In order to produce the high luminosity of the first-night light curve, a helium shell of about $0.1~M_{\odot}$ is required (panel (b) in Figure \ref{fig:EarlyLightCurveFittings}; refer to models 8A and 9A of \citealp{maeda18} for detailed model descriptions). The predicted timescale is, however, much longer than observed.

In the grid of the companion-ejecta interaction models with a range of binary configuration, a separation of $\sim$$2.0\times10^{12}$ cm between a companion star and the SN ejecta can roughly explain the prominent early excess while the early-excess timescale is significantly longer than that of SN~2020hvf (panel (c) in Figure \ref{fig:EarlyLightCurveFittings}; refer to model D2e12 of \citealp{maeda18} for details).

Synthesized light curves with a series of $^{56}$Ni distributions are shown in panel (d) of Figure \ref{fig:EarlyLightCurveFittings}. In these models, we add the surface $^{56}$Ni to our fiducial ejecta model. Specifically, the mass fraction of $^{56}$Ni is set as 50\% from the surface down to a specified velocity, taken to be either 10,000, 15,000, or 20,000 km s$^{-1}$; the corresponding $^{56}$Ni mass in the outermost region is 0.16, 0.05, or 0.01 $M_\odot$, respectively. For the latter two, the $^{56}$Ni distributions are indeed not monotonic and the major $^{56}$Ni core and the surface-$^{56}$Ni region are separated. For these three models, the early-phase light curves are computed with SNEC as in the same manner as with the CSM-ejecta interaction model. Despite the model sequence including the non-monotonic $^{56}$Ni distribution, we find that the light curve always shows a monotonically rising behavior unlike the observed one, and thus we conclude that the excess in the $^{56}$Ni mass in the outermost region would never explain the observed early-phase behavior.

\subsection{Overall Feature Modeling}

To interpret the overall light curve and spectral features, we have constructed a series of phenomenological ejecta models. The ejecta model construction follows the set up as described by \citet{maeda18}. The density structure is assumed to follow an exponential function in velocity, specified by the ejecta mass and the kinetic energy. The abundance stratification is then assumed to be divided into the electron-capture core, $^{56}$Ni-rich core, the Si-rich zone, then the O-rich layer from the inner to the outer region. No He-detonation layer is introduced for the present work. In summary, we have the kinetic energy and the masses of the four abundance zones as the input model parameters. Indirectly, the model parameters provide a required binding energy of the progenitor WD by subtracting the amount of the nuclear energy generation (specified by the composition structure) by the final kinetic energy adopted in the model.

The ejecta models are used as an input to model detailed radiation transfer simulations using HEIMDALL (Handling Emission In Multi-Dimension for spectrAL and Light curve calculations; \citealp{maeda14}). Note that the ejecta models have been input to the SNEC radiation-hydrodynamic simulations to study the CSM-ejecta interaction in Section 5.1. However, to study the overall light curve and spectral features, we directly use the ejecta model without CSM interaction. For the CSM properties we finally adopt (0.01 $M_\odot$ and $10^{13}$ cm), the energy created by the CSM interaction is quickly lost, and the density structure is changed only in the outermost region. Therefore, in the post-early-flash phase as a main interest of this section, the effect of the CSM interaction is largely negligible (more details will be given in a forthcoming paper; K. Maeda et al., in preparation). 

\begin{figure*}
\plotone{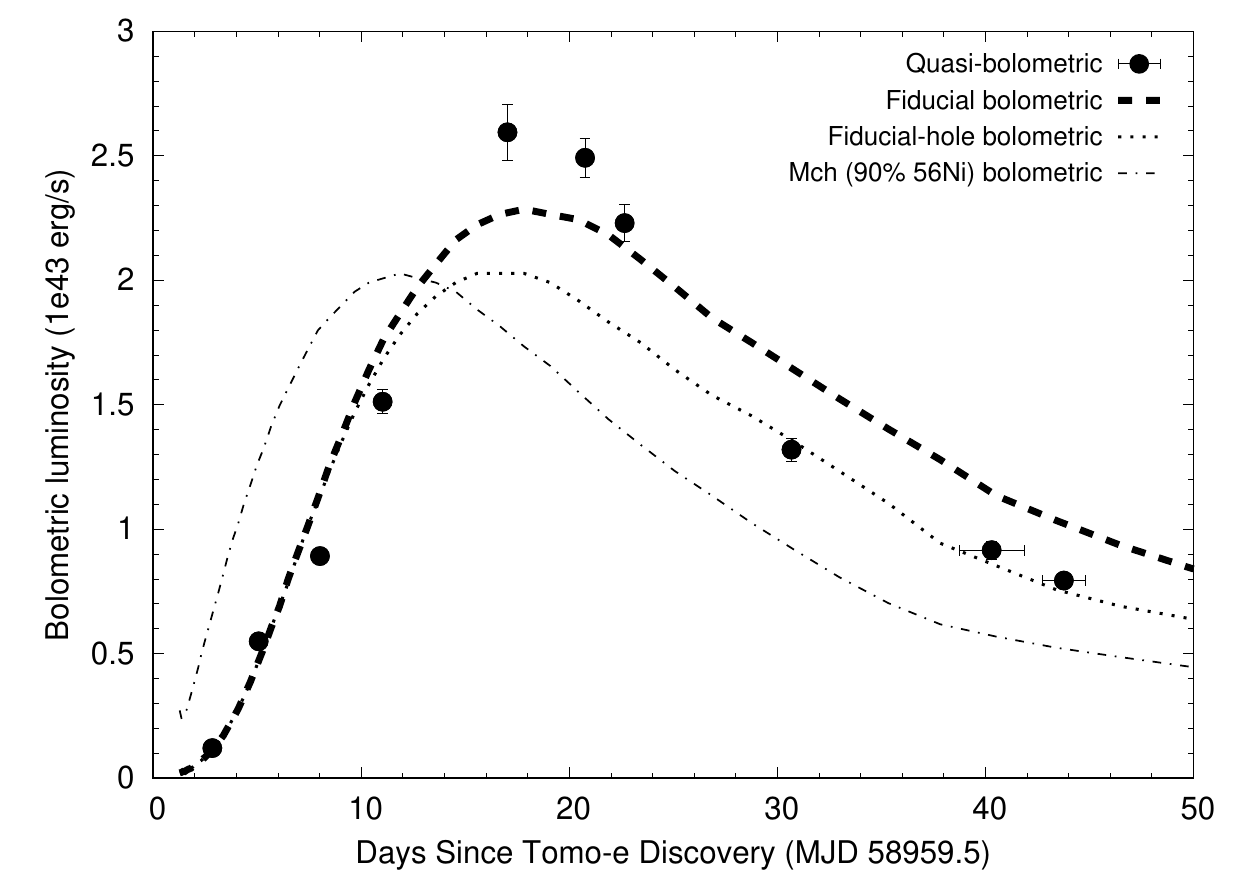}
\vspace{5pt}
\caption{The UVOIR pseudo bolometric light curve of SN~2020hvf and bolometric light curves synthesized by our fiducial super-$M\rm_{Ch}$ (dashed line), fiducial-hole (dotted line), and $1.26~M_{\odot}$ $^{56}$Ni $M\rm_{Ch}$ (dashed-dotted line) models, respectively.
\label{fig:Bolometric-LC}}
\end{figure*}

We first try to obtain a reasonable model sequence that can explain the post-early-flash phase. As the first trial we fix the ejecta mass as $1.4~M_\odot$ (hereafter ``$M\rm_{Ch}$ model"), but never find a model that explains the data; the model light curves always evolve too quickly for a reasonable range of the near-Chandrasekhar-mass WD binding energy given that nearly entire ejecta must be converted to $^{56}$Ni to reach the peak luminosity as high as observed, leading to a high ratio of the kinetic energy to the mass and thus short diffusion timescale. In Figure \ref{fig:Bolometric-LC}, a dashed-dotted line corresponding to the bolometric light curve of a $M\rm_{Ch}$ model for which all the ejecta are converted to the $^{56}$Ni-rich abundance (90\% $^{56}$Ni, 5\% stable Ni, 5\% stable Fe; i.e., a total $^{56}$Ni of $1.26~M_\odot$) evolves too quickly to be compatible with the slow-evolving light curve of SN~2020hvf.

The discrepancy demonstrates the difficulty in explaining the light curve of SN~2020hvf by variants of a $M\rm_{Ch}$ model---reducing the $^{56}$Ni mass will slow down the evolution speed but decrease the luminosity and never bring this specific model light curve to the characteristic slow and bright combination for the SN~2020hvf. We then increase the ejecta mass and find a reasonable solution for the ejecta mass of $\sim$2.1 $M_\odot$ and the kinetic energy of $1.4 \times 10^{51}$ erg. In our fiducial model we fix the mass of the $^{56}$Ni-rich zone to be 1.6 $M_\odot$ (1.44 $M_\odot$ of $^{56}$Ni assuming a typical fraction of $^{56}$Ni as 90\%). By also varying the Si- and O-rich zones, we adopt 0.3 and 0.2 $M_\odot$ for these zones, respectively.

\begin{figure*}
\plotone{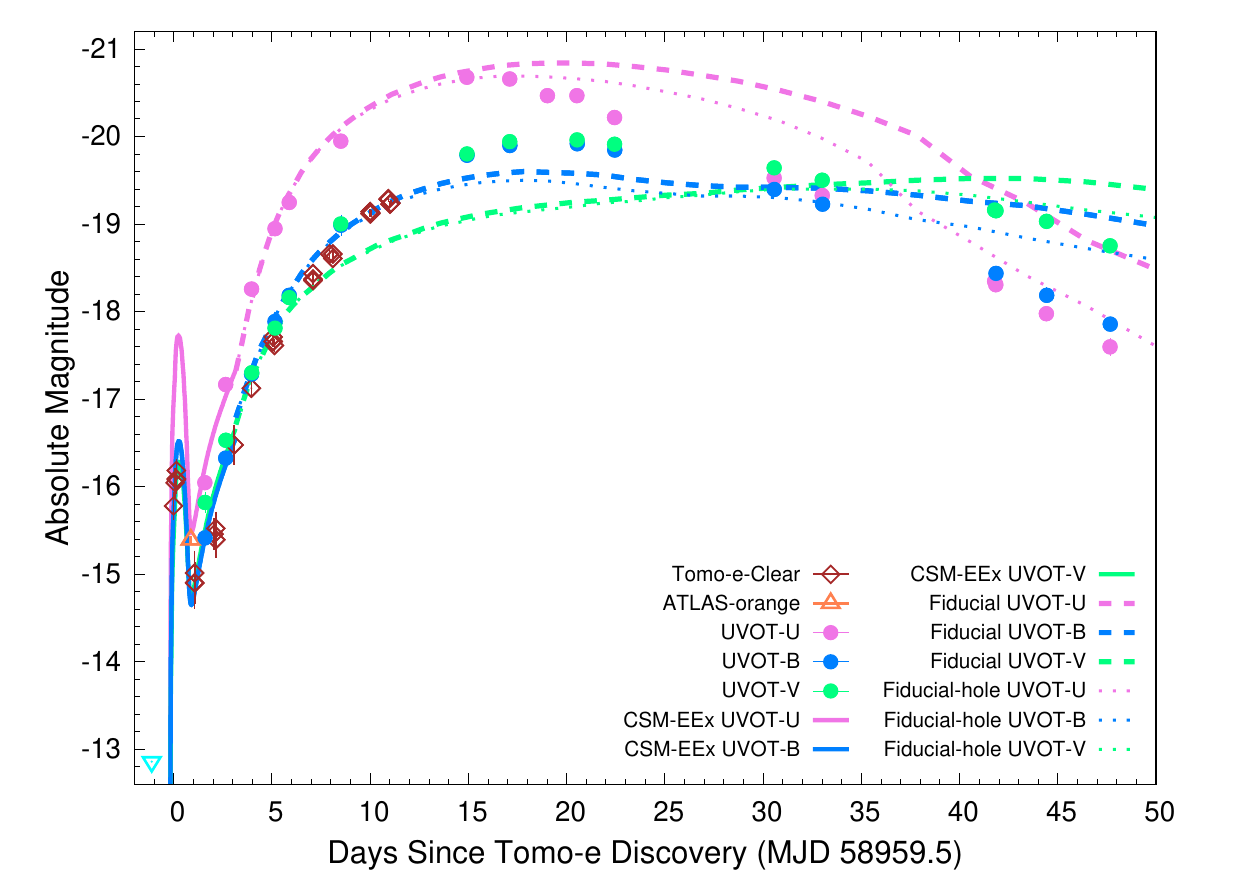}
\vspace{0pt}
\caption{Synthesized $U, B, V$ light curves of SN~2020hvf. Thick solid lines correspond to the best-fit CSM-interaction model (i.e., CSM mass of 0.01 $M_{\odot}$ and a characteristic radius of $1.0\times10^{13}$ cm) for explaining the light-curve behavior in the first few days. Dashed lines are light curves (``fiducial") purely powered by radioactive decay of 1.44 $M_\odot$ $^{56}$Ni as an explosion of a super-$M\rm_{Ch}$ WD (2.1 $M_{\odot}$ and $1.4 \times 10^{51}$ erg). Dotted lines assume that the innermost 0.3 $M_\odot$ $^{56}$Ni-rich core of the fiducial model is replaced by an electron-capture core (``fiducial-hole"), which gives a slightly better fitting in the declining phase; this model has a total 1.2 $M_\odot$ of $^{56}$Ni (Section 6.2). The best-fit CSM-interaction model indicates that the SN~2020hvf explodes at about 0.2 day before the Tomo-e discovery.
\label{fig:Overall_LC_Fitting}}
\end{figure*}

The light curve of this ejecta model is shown as the fiducial model in Figure \ref{fig:Overall_LC_Fitting} (dashed lines). The spectral evolution is shown in Figure \ref{fig:sn2020hvf_spec_evo} (red). Given that our main focus here is to constrain the general ejecta properties and test the super-$M\rm_{Ch}$ WD hypothesis for SN~2020hvf, we have not tuned the model to obtain a better fit (see below for further discussion). Rather, the model spectra are shown for a demonstration purposes, as a sanity check to show that the model constructed based on the light-curve properties can also provide spectra largely consistent with the observed ones. As shown in Figure \ref{fig:Overall_LC_Fitting}, the overall light curve is explained reasonably well, and the key features in the spectra (Figure \ref{fig:sn2020hvf_spec_evo}) are also explained without fine-tuning. Specifically, the evolution of the Si $\rm _{II}$ velocity is roughly reproduced. The relatively high velocity of the Si $\rm _{II}$ line at maximum brightness is found to be consistent with the super-$M\rm_{Ch}$ WD explosion model. It is interesting to note that the model naturally produces the very high velocity in the initial phase and it quickly decreases to normal velocity without introducing additional modification to the standard exponential density structure. This is indeed a challenge when explaining similar high-velocity features in normal SNe Ia \citep{gerardy04,tanaka08}. For the present model, the combination of the high luminosity and the small photospheric radius during the early-phase results in sufficient amounts of single-ionized Si in the outermost layer, leading to the formation of the high-velocity feature in the initial phase.

Another key spectral feature of SN~2020hvf as a carbon-rich overluminous SN Ia is the existence of the C $\rm _{II}$ absorption. We, however, did not attempt to explain this feature; the ejecta model indeed does not contain carbon (i.e., there is no carbon introduced in the outermost O layer). 
The formation of C $\rm _{II}$ is not well understood in general, not only in the overluminous SNe Ia but also in normal SNe Ia. 
In addition, the distance uncertainty could limit the detailed modeling of relatively weak features like C $\rm _{II}$, as the line formation can be very sensitive to the model temperature and thus the intrinsic luminosity. More details will be discussed in a separate paper (K. Maeda et al., in preparation).

\section{Discussion and Conclusions}

\subsection{The Early Excess of SN~2020hvf}

Among the four early-excess scenarios, i.e., (a) CSM-ejecta interaction, (b) He-det, (c) companion-ejecta interaction, and (d) surface-$^{56}$Ni-decay scenarios, the CSM-ejecta interaction model can nicely reproduce the earliest phase light curve, while the other three have difficulty in explaining the combination of the luminosity and the duration (or the characteristic evolution). In this section, we discuss details of this issue.

The strong early emission and high brightness in UV/NUV wavelengths soon after the early emission of SN~2020hvf conflict with the prediction of the He-det scenario. This is because, in the He-det scenario, a considerable amount of titanium and iron-group element absorptions below $\sim$4500 \text{\AA} will be generated in order to explain the high luminosity of the early flash of SN~2020hvf \citep{JJA2017,maeda18}, which results in a much redder $B-V$ color in the next few days after the early excess (panel (b) in Figure \ref{fig:EarlyLightCurveFittings}). In addition, theoretically, the He-det-induced explosion mechanisms (e.g., the double-detonation scenario; \citealp{livne90,woosley94}) predict that the onset of the SN explosion occurs before the mass of the primary WD approaching the Chandrasekhar limit, leading to a subluminous or normal-brightness SN Ia \citep{fink10,kromer10,pakmor13,tanikawa19}.

The surface-$^{56}$Ni-decay-induced early excess requires an extended $^{56}$Ni distribution and a considerable amount of $^{56}$Ni has to be formed at the outermost region of the SN ejecta \citep{piro16,magee20}. To explain such a bright pulse-like early excess isolated from the major light curve with the surface-$^{56}$Ni-decay scenario, most of the $^{56}$Ni should be confined to the inner layers while a considerable amount of $^{56}$Ni has to be formed at the surface of the ejecta. However, such a configuration is hard to realize for SNe Ia; the ``bulk" $^{56}$Ni distribution is already extended to a high velocity and there is no room for having a sufficient amount of materials between the $^{56}$Ni core and the surface (which is required to produce the ``dip" after the earliest phase). Our simulations (panel (d) in Figure \ref{fig:EarlyLightCurveFittings}) do confirm that in order to reach the brightness of the early flash via the surface-$^{56}$Ni decay, we should see a bright but bump-like early excess \citep{JJA2018,JJA2020}, which contradicts the bright ``pulse-like" early excess discovered in SN~2020hvf. Furthermore, unlike 91T/99aa-like overluminous SNe Ia, prominent IME adsorptions were found in the early-phase spectra of SN~2020hvf (Figure \ref{fig:sn2020hvf_spec_evo}), suggesting a less efficient detonation approaching the outermost region of the SN ejecta.

A series of analytical and numerical simulations were performed to test whether or not the two interaction-induced early-excess scenarios can generate the bright pulse-like early excess as shown in Section 5.1. Although the luminosity of the companion-interaction-induced early excess can be comparable to that of SN~2020hvf as shown in panel (c) in Figure \ref{fig:EarlyLightCurveFittings}, its timescale is much longer than the observed early excess. For example, our model with $5.0\times10^{11}$ cm separation between the SN center and the companion provides an optical luminosity already below the observation while it still has a longer timescale. Given that a smaller separation (and thus companion radius) leads to a fainter and faster early-excess emission, to have an even shorter timescale, we will need to further reduce the separation/radius, which will result in even fainter luminosity. In addition, within the single-degenerate scenario, we would not expect a configuration where the companion star is much smaller in radius than the models here. In contrast, the early excess caused by the CSM-ejecta interaction evolves generally faster due to a shorter characteristic timescale of the evolution in the bolometric light curve of the CSM-induced early excess \citep{maeda18}, which nicely fits the prominent pulse-like early excess (panel (a), Figure \ref{fig:EarlyLightCurveFittings}) for a CSM mass of $0.01~M_{\odot}$ and a characteristic (outer edge) radius of $1.0\times10^{13}$ cm. The nondetection of H $\rm _{II}$ and He $\rm _{II}$ features in spectra of SN~2020hvf suggests that the CSM is likely carbon-rich, originating from either debris of a disrupted CO WD or surface materials of a spin-down super-$M\rm_{Ch}$ CO WD.

\subsection{The Progenitor of SN~2020hvf and Implications of the Origin(s) of Carbon-rich Overluminous SNe Ia}

Our understanding on the origin(s) of carbon-rich overluminous SNe Ia has been limited mainly due to the uncertainty of the energy source that contributes to the ultra-high luminosity \citep{taubenberger17}. Although a significant amount of $^{56}$Ni from a super-$M\rm_{Ch}$ WD explosion can intuitively explain the high luminosity of carbon-rich overluminous SNe Ia, other scenarios have been proposed to avoid the abnormal super-$M\rm_{Ch}$ WD progenitor. 

One prevailing scenario is the interaction between a very extended, massive CSM envelope and the ejecta from a normal WD explosion \citep{hicken07,scalzo10,taubenberger11,taubenberger13}. Thanks to the stringent constraint given by early-excess modeling, the confined CSM distribution derived from the early emission rules out the possibility of a CSM-interaction origin of the high luminosity of SN~2020hvf; if the peak luminosity is powered by the CSM interaction, that must dilute any short-timescale variability in the rising phase, irrespective of the origin of such a flash.

Another scenario that might explain overluminous SNe Ia with a smaller amount of $^{56}$Ni is an asymmetric $^{56}$Ni distribution after the SN explosion \citep{hillebrandt07,leung21}. However, the asymmetric $^{56}$Ni distribution is unlikely to explain SN~2020hvf. If the main power source is the asymmetrically distributed $^{56}$Ni, this cannot create the pulse-like early excess of SN~2020hvf within the same context and, instead, it is possible that a broader rising-phase light curve with a bump-like early excess commonly shown in 91T/99aa-like overluminous SNe Ia can be expected (i.e., similar to the surface-$^{56}$Ni-decay scenario discussed above). Thus, a different mechanism is required to explain the early-excess feature of SN~2020hvf. Also, there is no viewing-angle effect shown among the carbon-rich overluminous SNe Ia. For example, we do not see correlations among the SN luminosity, the strengths of intermediate-mass element absorptions, and the light-curve evolving speed \citep{ashall21}.

The main drawback of the super-$M\rm_{Ch}$ WD progenitor model is the poor fitting of the $V$-band light curve around the peak and overfitting in blue wavelengths in the declining phase (Figure \ref{fig:Overall_LC_Fitting}). We further investigate model light curves by reducing the $^{56}$Ni amount in the inner region of the ejecta, and the result shows that the overall light curve cannot be promisingly improved by reducing $\sim$17$\%$ of $^{56}$Ni (the ``fiducial-hole" model; dotted lines in Figure \ref{fig:Overall_LC_Fitting}). Given the unpromising output from the less massive $^{56}$Ni model and a reasonable fitting of the overall bolometric light curves, especially around the peak, via our fiducial super-$M\rm_{Ch}$ WD model (Figure \ref{fig:Bolometric-LC}), we suggest that the super-$M\rm_{Ch}$ WD model is a promising scenario and the models shown here are thus regarded as defining a range of the super-$M\rm_{Ch}$ SN Ia light curves. However, additional hydrodynamic and radiation transfer simulations are required for further understanding of the multiband light-curve behavior of the SN~2020hvf.

For the super-$M\rm_{Ch}$ WD progenitor, both the mechanism and progenitor system leading to such an abnormal condition are under debate. In the early 2000s, theoretical studies have predicted that a WD can reach 1.2--2.7~$M_{\odot}$ by increasing its spin speed through the single-degenerate channel  \citep{langer00,yoon05,hachisu12,benvenuto15}. Recently, a series of smoothed particle hydrodynamic simulations have been performed to investigate if a merger of two WDs via the double-degenerate channel can finally lead to SN Ia events. The results show a large uncertainty in producing SNe Ia without a foregoing He-shell detonation that triggers a central carbon ignition of the primary WD \citep{dan11,dan12,dan14,schwab12,shen12,tanikawa15}. Therefore, it is still an open question as to whether typical SNe Ia can be produced through the merger of two CO WDs with a total mass above the Chandrasekhar limit. However, the simulations suggest that ultraluminous SNe Ia can be expected once the total mass of the binary CO WD system is significantly beyond the Chandrasekhar limit (i.e., a merger of two massive CO WDs; \citealp{dan12,dan14,zhu13}). As such massive systems will be rare in the universe, this might explain the extremely low event rate of the carbon-rich overluminous SNe Ia.

A further question is how the (carbon-abundant) CSM envelope can be formed. We expect that the existence and the nature of the CSM can distinguish between the single- and double-degenerate channels (among other possible scenarios) leading to the formation of a super-massive WD and a subsequent explosion. If the massive WD, beyond the Chandrasekhar limit for a nonrotating configuration, was created by accretion from a nondegenerate companion star, the WD likely evolves through the critical rotation where the centrifugal force is marginally balanced with the gravity near the surface \citep{uenishi03,yoon05}. This requires the angular momentum redistribution, and a fraction of the accretion materials may finally form a disk or torus around the WD. Alternatively, but through a similar mechanism, a binary WD merger likely creates a massive torus as an immediate outcome of the merger, which then evolves into a hot envelope due to further angular momentum redistribution, probably through the magnetic field \citep{schwab12,shen12}. In the former case, the composition could be either H/He-rich or carbon-rich, while in the latter, it is expected to be carbon-rich. Future studies, both on placing a stronger constraint on the CSM composition and on the detailed evolution calculations leading to the super-$M\rm_{Ch}$ WD, will therefore be necessary.

A remaining question is whether all carbon-rich overluminous SNe Ia discovered so far have the same origin. The required binding energy of 1.6 $\times~10^{51}$ erg of the WD progenitor of SN~2020hvf (2.1 $M_{\odot}$) is within the theoretical model sequence with a range of the central density \citep{yoon05}. The WD mass is very close to the upper limit of the theoretical prediction (2.1 $M_{\odot}$ in the model sequence by \citealp{yoon05}). If the $^{56}$Ni mass is larger for a more massive WD, it may require an even more massive WD in order to generate more luminous carbon-rich overluminous SNe Ia such as SN~2009dc. Furthermore, the low velocity seen in some carbon-rich overluminous SNe Ia would require a higher binding energy, which will raise another challenge; indeed, most of the previous examples of carbon-rich overluminous SNe Ia show observed ejecta velocities significantly lower than the prediction \citep{maeda09b,hachinger12}.

In contrast, the high ejecta velocity of SN~2020hvf perfectly matches the prediction of exploding a super-$M\rm_{Ch}$ WD. Interestingly, the other high-velocity carbon-rich overluminous SN Ia, ASASSN-15pz, likely shows a pulse-like early emission, though a large photometric uncertainty makes it difficult to give a robust judgement \citep{chen19}. The detection of (probable) pulse-like early excess in both high-velocity carbon-rich overluminous SNe Ia may indicate a specific pathway of forming the super-$M\rm_{Ch}$ WD. It also suggests a different origin of the 20hvf-like overluminous SNe Ia compared to the typical carbon-rich overluminous (i.e., 03fg/09dc-like) SNe Ia. To figure out the possible multiple origins of carbon-rich overluminous SNe Ia, the early-phase intranight photometry will play an irreplaceable role in enlightening us about the CSM distribution, explosion mechanism, and progenitor system of the carbon-rich overluminous SNe Ia. With the ongoing high-cadence transient surveys such as the Zwicky Transient Facility (ZTF; \citealp{bellm19a,bellm19b,graham19}), the Tomo-e Gozen high-cadence survey, and the MUlti-band Subaru Survey for Early-phase Supernovae (MUSSES; \citealp{JJA2017}), systematical studies of the early-phase carbon-rich overluminous SNe Ia can be achieved in the coming years.

\subsection{Summary}

In this Letter we report the observation of a carbon-rich, overluminous SN Ia, SN~2020hvf, discovered by the Tomo-e Gozen camera within about 0.2 day of the explosion. The prominent pulse-like early flash is explained almost exclusively by an interaction between the supernova ejecta and $\sim$0.01 $M_{\odot}$ CSM extending to a distance of $\sim$$10^{13}~\text{cm}$. On the basis of the CSM-induced prompt early excess, the overluminous light curve, and the high ejecta velocity of the SN~2020hvf, we suggest that the SN~2020hvf may originate from a thermonuclear explosion of a super-$M\rm_{Ch}$ WD. A detailed comparison with models is limited, however, by several factors; there is no detailed explosion model developed so far for the super-$M\rm_{Ch}$ WD explosion. The distance uncertainty affects the derived ejecta density and temperature, which substantially affect the light curve and spectral formation. Given the large diversity of carbon-rich overluminous SNe Ia discovered so far, systematical investigations from the very early phase via high-cadence time-domain surveys are required to figure out the origin(s) of this extreme SN Ia subclass.

\vspace{30pt}

We thank the anonymous referee for helpful comments and suggestions. This work has been supported by the Japan Society for the Promotion of Science (JSPS) KAKENHI grant JP18J12714 and JP19K23456 (J.J.); JP16H01087 and JP18H04342 (J.J. and M.D.); JP18H05223 (J.J., K.M., M.D., and T.S.); JP20H00174 and JP20H04737 (K.M.); JP21K13959 (M.K.); JP15H02082, JP16H06341, and JP16K05287 (T.S.); JP17H06363 (M.T. and S.S.); JP19H00694, JP20H00158, and JP20H00179 (M.T.); JP16H02158 (N.T. and S.S.); JP17K05382, JP20K04024, and JP21H04499 (K.N.); JP21H04491 (S.S.); JP18K13599 (R.O.); JP20K14521(K.I.); JP18H01261 (S.O.) and JP18H04575.

This work is supported in part by the Research Center for the Early Universe (RESCEU) of the School of Science at the University of Tokyo, the World Premier International Research Center Initiative (WPI), MEXT, Japan, and the Optical and Near-Infrared Astronomy Inter-University Cooperation Program, supported by the MEXT of Japan.
Based on observations made with the Thai Robotic Telescopes under program IDs TRTC07C\_014, TRTC07D\_003 and TRTC08B\_011, which are operated by the National Astronomical Research Institute of Thailand (Public Organization).

\facilities{Kiso:1.05 m, Seimei, Kanata, Swift, DSO17, RRRT, TRT.}

\software{Astropy \citep{astropy13,astropy18}, IRAF, GALFIT\citep{galfit10}, HEAsoft\citep{heasoft96}, SNEC\citep{morozova15}.}

\bibliographystyle{aasjournal}
\bibliography{main.bbl}

\appendix

\startlongtable
\begin{deluxetable*}{ccccccccccc}
\movetabledown=3000mm
\tablenum{1}
\tablecaption{Ground-based Optical/NIR Imaging observations of SN~2020hvf\label{tab:Ground Photometry Data}}
\tablewidth{0pt}
\tablehead{
\colhead{UT Date} & \colhead{Phase$^*$} & \colhead{Tel$^{\dagger}$} & \colhead{$Clear$} & \colhead{$B$} & \colhead{$V$} & \colhead{$R$} & \colhead{$I$} & \colhead{$J$} & \colhead{$H$} & \colhead{$K_{s}$} 
}
\startdata
Apr 20.5019 & -19.1913 & 1 & 16.77 (16) & -- & -- & -- & -- & -- & -- & -- \\
Apr 20.5830 & -19.1107 & 1 & 16.51 (17) & -- & -- & -- & -- & -- & -- & -- \\
Apr 20.6426 & -19.0514 & 1 & 16.37 (13) & -- & -- & -- & -- & -- & -- & -- \\
Apr 20.6439 & -19.0501 & 1 & 16.47 (16) & -- & -- & -- & -- & -- & -- & -- \\
Apr 20.6441 & -19.0499 & 1 & 16.46 (15) & -- & -- & -- & -- & -- & -- & -- \\
Apr 21.5771 & -18.1223 & 1 & 17.65 (30) & -- & -- & -- & -- & -- & -- & -- \\
Apr 21.5773 & -18.1221 & 1 & 17.53 (25) & -- & -- & -- & -- & -- & -- & -- \\
Apr 21.5809 & -18.1185 & 1 & 17.65 (24) & -- & -- & -- & -- & -- & -- & -- \\
Apr 22.5735 & -17.1317 & 1 & 17.09 (18) & -- & -- & -- & -- & -- & -- & -- \\
Apr 22.6559 & -17.0497 & 1 & 17.03 (19) & -- & -- & -- & -- & -- & -- & -- \\
Apr 22.6561 & -17.0495 & 1 & 17.16 (20) & -- & -- & -- & -- & -- & -- & -- \\
Apr 23.55 & -16.16 & 2 & -- & 16.39 (04) & 16.14 (02) & 15.90 (03) & 15.93 (05) & -- & -- & -- \\
Apr 23.5969 & -16.1142 & 1 & 16.07 (23) & -- & -- & -- & -- & -- & -- & -- \\
Apr 24.4829 & -15.2333 & 1 & 15.42 (09) & -- & -- & -- & -- & -- & -- & -- \\
Apr 25.51 & -14.21 & 2 & -- & 15.07 (02) & 14.85 (02) & 14.67 (03) & 14.67 (04) & -- & -- & -- \\
Apr 25.55 & -14.17 & 3 & -- & -- & 14.67 (03) & 14.53 (05) & 14.53 (05) & 14.82 (16) & 14.40 (18) & 14.67 (18) \\
Apr 25.5883 & -14.1343 & 1 & 14.89 (07) & -- & -- & -- & -- & -- & -- & -- \\
Apr 25.5885 & -14.1341 & 1 & 14.84 (06) & -- & -- & -- & -- & -- & -- & -- \\
Apr 25.5949 & -14.1277 & 1 & 14.89 (07) & -- & -- & -- & -- & -- & -- & -- \\
Apr 25.6630 & -14.0600 & 1 & 14.94 (07) & -- & -- & -- & -- & -- & -- & -- \\
Apr 27.6367 & -12.0977 & 1 & 14.20 (05) & -- & -- & -- & -- & -- & -- & -- \\
Apr 27.6369 & -12.0975 & 1 & 14.18 (05) & -- & -- & -- & -- & -- & -- & -- \\
Apr 27.6380 & -12.0964 & 1 & 14.12 (05) & -- & -- & -- & -- & -- & -- & -- \\
Apr 28.4756 & -11.2637 & 1 & 13.89 (04) & -- & -- & -- & -- & -- & -- & -- \\
Apr 28.52 & -11.22 & 2 & -- & 13.89 (04) & 13.81 (02) & 13.74 (03) & 13.76 (04) & -- & -- & -- \\
Apr 28.61 & -11.13 & 3 & -- & -- & 13.67 (03) & 13.64 (03) & 13.68 (05) & 13.43 (09) & 13.46 (12) & 13.28 (06) \\
Apr 28.6486 & -11.0917 & 1 & 13.89 (04) & -- & -- & -- & -- & -- & -- & -- \\
Apr 28.6488 & -11.0915 & 1 & 13.94 (04) & -- & -- & -- & -- & -- & -- & -- \\
Apr 30.5578 & -9.1935 & 1 & 13.41 (03) & -- & -- & -- & -- & -- & -- & -- \\
Apr 30.5591 & -9.1922 & 1 & 13.43 (03) & -- & -- & -- & -- & -- & -- & -- \\
Apr 30.5593 & -9.1920 & 1 & 13.41 (03) & -- & -- & -- & -- & -- & -- & -- \\
Apr 30.5872 & -9.1643 & 1 & 13.43 (03) & -- & -- & -- & -- & -- & -- & -- \\
Apr 30.5885 & -9.1630 & 1 & 13.43 (03) & -- & -- & -- & -- & -- & -- & -- \\
Apr 30.5887 & -9.1628 & 1 & 13.42 (03) & -- & -- & -- & -- & -- & -- & -- \\
May 01.4900 & -8.2667 & 1 & 13.27 (03) & -- & -- & -- & -- & -- & -- & -- \\
May 01.4910 & -8.2657 & 1 & 13.26 (03) & -- & -- & -- & -- & -- & -- & -- \\
May 01.4921 & -8.2646 & 1 & 13.27 (03) & -- & -- & -- & -- & -- & -- & -- \\
May 01.58 & -8.18 & 2 & -- & 13.36 (02) & 13.19 (02) & 13.24 (03) & -- & -- & -- & -- \\
May 01.5756 & -8.1816 & 1 & 13.32 (03) & -- & -- & -- & -- & -- & -- & -- \\
May 01.61 & -8.14 & 3 & -- & -- & 13.09 (03) & 13.14 (03) & 13.20 (05) & 12.94 (12) & 13.02 (03) & 12.85 (15) \\
May 01.6240 & -8.1334 & 1 & 13.31 (03) & -- & -- & -- & -- & -- & -- & -- \\
May 02.19 & -7.57 & 4 & -- & -- & 13.01 (05) & 13.05 (05) & 13.26 (05) & -- & -- & -- \\
May 03.05 & -6.72 & 5 & -- & 12.88 (06) & 12.97 (03) & 13.02 (05) & -- & -- & -- & -- \\
May 03.11 & -6.66 & 4 & -- & -- & 12.95 (05) & 12.98 (05) & -- & -- & -- & -- \\
May 05.12 & -4.65 & 5 & -- & 12.65 (06) & 12.78 (03) & 12.87 (05) & -- & -- & -- & -- \\
May 05.20 & -4.58 & 6 & -- & 12.78 (07) & 12.83 (05) & 12.83 (05) & -- & -- & -- & -- \\
May 06.66 & -3.13 & 2 & -- & -- & 12.73 (02) & 12.92 (03) & 12.99 (04) & -- & -- & -- \\
May 07.09 & -2.70 & 4 & -- & -- & 12.66 (04) & 12.68 (05) & 13.03 (05) & -- & -- & -- \\
May 07.55 & -2.24 & 2 & -- & 12.90 (03) & 12.68 (02) & 12.87 (03) & 12.93 (04) & -- & -- & -- \\
May 07.66 & -2.13 & 3 & -- & -- & 12.73 (05) & 12.86 (05) & 12.99 (07) & 12.25 (08) & 12.21 (14) & 11.88 (36) \\
May 08.05 & -1.75 & 5 & -- & 12.64 (06) & 12.65 (03) & 12.73 (05) & -- & -- & -- & -- \\
May 08.09 & -1.70 & 4 & -- & -- & 12.63 (05) & 12.67 (05) & 13.01 0.05 & -- & -- & -- \\
May 08.17 & -1.62 & 4 & -- & -- & 12.63 (05) & -- & -- & -- & -- & -- \\
May 09.66 & -0.14 & 6 & -- & -- & 12.61 (05) & -- & -- & -- & -- & -- \\
May 10.07 & 0.27 & 5 & -- & 12.60 (06) & 12.61 (03) & 12.71 (05) & -- & -- & -- & -- \\
May 10.11 & 0.30 & 4 & -- & -- & 12.61 (05) & 12.62 (05) & 12.99 (05) & -- & -- & -- \\
May 11.07 & 1.26 & 5 & -- & 12.65 (06) & 12.64 (03) & 12.67 (05) & -- & -- & -- & -- \\
May 11.10 & 1.29 & 4 & -- & -- & 12.62 (05) & 12.60 (05) & 12.96 (05) & -- & -- & -- \\
May 11.54 & 1.73 & 2 & -- & 12.91 (03) & 12.62 (06) & 12.78 (03) & 12.91 (05) & -- & -- & -- \\
May 11.62 & 1.81 & 3 & -- & -- & 12.61 (03) & 12.70 (04) & 12.85 (05) & 12.27 (09) & 12.27 (04) & 12.06 (02) \\
May 12.10 & 2.28 & 4 & -- & -- & 12.64 (04) & 12.60 (05) & 12.97 (05) & -- & -- & -- \\
May 12.55 & 2.73 & 3 & -- & -- & 12.60 (02) & 12.69 (03) & 12.82 (05) & 12.64 (20) & 12.56 (07) & 12.25 (08) \\
May 13.52 & 3.69 & 2 & -- & 12.98 (02) & 12.70 (02) & 12.77 (03) & 12.92 (04) & -- & -- & -- \\
May 14.07 & 4.24 & 5 & -- & 12.78 (06) & 12.69 (03) & 12.71 (05) & -- & -- & -- & -- \\
May 15.14 & 5.31 & 4 & -- & -- & 12.72 (04) & 12.66 (05) & 13.00 (05) & -- & -- & -- \\
May 16.11 & 6.27 & 5 & -- & 12.85 (06) & -- & 12.75 (05) & -- & -- & -- & -- \\
May 16.16 & 6.32 & 4 & -- & -- & -- & 12.71 (05) & 13.00 (05) & -- & -- & -- \\
May 17.16 & 7.31 & 4 & -- & -- & -- & 12.74 (05) & 13.06 (05) & -- & -- & -- \\
May 17.38 & 7.53 & 6 & -- & 12.78 (07) & -- & 12.78 (05) & 12.97 (06) & -- & -- & -- \\
May 19.51 & 9.65 & 2 & -- & 13.32 (03) & 12.92 (02) & 13.00 (03) & 13.11 (04) & -- & -- & -- \\
May 20.49 & 10.62 & 2 & -- & 13.40 (03) & 12.98 (02) & 13.06 (03) & 13.16 (04) & -- & -- & -- \\
May 21.58 & 11.71 & 2 & -- & 13.45 (03) & 13.03 (03) & 13.08 (03) & 13.17 (05) & -- & -- & -- \\
May 22.49 & 12.61 & 2 & -- & 13.53 (03) & 13.06 (02) & 13.11 (03) & 13.14 (04) & -- & -- & -- \\
May 22.59 & 12.71 & 3 & -- & -- & -- & -- & -- & 13.46 (11) & 12.57 (13) & 12.46 (10) \\
May 24.08 & 14.19 & 5 & -- & 13.40 (06) & 13.13 (04) & 13.09 (05) & -- & -- & -- & -- \\
May 24.53 & 14.64 & 2 & -- & 13.68 (02) & 13.15 (02) & 13.12 (03) & 13.10 (04) & -- & -- & -- \\
May 28.53 & 18.62 & 2 & -- & 14.05 (02) & 13.30 (02) & 13.17 (03) & 13.05 (04) & -- & -- & -- \\
May 29.47 & 19.55 & 2 & -- & 14.14 (02) & 13.32 (02) & 13.17 (03) & 13.02 (04) & -- & -- & -- \\
May 29.53 & 19.61 & 3 & -- & -- & -- & -- & -- & 13.32 (03) & 12.61 (05) & 12.48 (08) \\
May 31.10 & 21.17 & 5 & -- & 14.10 (06) & -- & 13.17 (05) & -- & -- & -- & -- \\
Jun 01.09 & 22.15 & 5 & -- & 14.16 (07) & 13.58 (06) & 13.21 (05) & -- & -- & -- & -- \\
Jun 02.09 & 23.15 & 5 & -- & 14.30 (06) & 13.58 (06) & 13.22 (05) & -- & -- & -- & -- \\
Jun 02.49 & 23.55 & 2 & -- & 14.63 (03) & 13.52 (02) & 13.29 (03) & 13.04 (04) & -- & -- & -- \\
Jun 03.07 & 24.13 & 5 & -- & 14.47 (07) & -- & 13.27 (05) & -- & -- & -- & -- \\
Jun 03.11 & 24.17 & 4 & -- & -- & -- & -- & 13.14 (05) & -- & -- & -- \\
Jun 04.55 & 25.59 & 3 & -- & -- & -- & -- & -- & 13.16 (07) & 12.68 (03) & 12.44 (09) \\
Jun 08.09 & 29.11 & 5 & -- & 14.93 (06) & -- & 13.56 (05) & -- & -- & -- & -- \\
Jun 09.09 & 30.11 & 5 & -- & 14.97 (08) & -- & 13.62 (07) & -- & -- & -- & -- \\
Jun 09.49 & 30.51 & 3 & -- & -- & 13.80 (03) & 13.51 (04) & 13.08 (05) & -- & -- & -- \\
Jun 12.11 & 33.12 & 4 & -- & -- & -- & 13.67 (05) & -- & -- & -- & -- \\
Jun 13.08 & 34.08 & 4 & -- & -- & 14.18 (06) & -- & 13.42 (05) & -- & -- & -- \\
Jun 15.54 & 36.52 & 2 & -- & 15.48 (03) & 14.16 (02) & 13.97 (03) & 13.47 (04) & -- & -- & -- \\
Jun 15.57 & 36.55 & 3 & -- & -- & 14.13 (02) & 13.90 (04) & -- & -- & -- & -- \\
Jun 16.52 & 37.50 & 2 & -- & -- & 14.17 (02) & 14.01 (03) & 13.39 (04) & -- & -- & -- \\
Jun 19.48 & 40.44 & 2 & -- & 15.55 (02) & 14.27 (02) & 14.12 (03) & 13.66 (04) & -- & -- & -- \\
Jun 19.52 & 40.48 & 3 & -- & -- & 14.26 (03) & 14.09 (04) & 13.60 (06) & 13.76 (10) & 12.56 (25) & 13.00 (18) \\
Jun 21.48 & 42.43 & 2 & -- & 15.58 (03) & 14.33 (02) & 14.20 (03) & 13.72 (04) & -- & -- & -- \\
Jun 23.51 & 44.45 & 3 & -- & -- & 14.31 (03) & 14.21 (04) & 13.75 (05) & 14.01 (18) & 12.86 (11) & -- \\
Jun 24.09 & 45.02 & 4 & -- & -- & 14.48 (05) & -- & -- & -- & -- & -- \\
Jun 27.07 & 47.99 & 4 & -- & -- & 14.69 (05) & 14.21 (06) & -- & -- & -- & -- \\
Jun 27.07 & 47.99 & 5 & -- & 15.49 (18) & 14.63 (09) & 14.28 (06) & -- & -- & -- & -- \\
Jul 01.51 & 52.40 & 2 & -- & 15.61 (04) & 14.59 (04) & 14.59 (03) & 14.11 (04) & -- & -- & -- \\
Jul 02.38 & 53.26 & 6 & -- & 15.51 (08) & -- & 14.49 (06) & 14.35 (06) & -- & -- & -- \\
Jul 04.38 & 55.26 & 6 & -- & 15.50 (09) & 14.79 (05) & 14.57 (06) & 14.43 (07) & -- & -- & -- \\
Jul 05.36 & 56.23 & 6 & -- & 15.60 (08) & -- & -- & -- & -- & -- & -- \\
Jul 15.38 & 66.19 & 6 & -- & 15.79 (09) & -- & 14.96 (07) & 14.86 (08) & -- & -- & -- \\
Jul 16.36 & 67.17 & 6 & -- & -- & 15.05 (05) & -- & -- & -- & -- & -- \\
Jul 16.47 & 67.27 & 2 & -- & 15.77 (02) & 14.98 (03) & 15.07 (03) & 14.66 (04) & -- & -- & -- \\
Jul 24.43 & 75.19 & 6 & -- & -- & 15.39 (10) & -- & -- & -- & -- & -- \\
\hline
\enddata
\tablecomments{\\
The Tomo-e nonfilter photometries ($Clear$) are in the AB system. The magnitudes in other bands are in the Vega system. Numbers in parentheses correspond to 1$\sigma$ statistical uncertainties in units of 1/100 mag (absolute flux calibration error is not included for the Tomo-e
photometries). The Galactic extinction ($E(B-V)\rm_{MW}$ = 0.0356 mag) has been corrected.\\
$^*$ Days (rest frame) relative to the estimated date of the $B$-band maximum, 2020 May 9.8047.\\
$^{\dagger}$ Telescope codes: 1: Kiso Schmidt/Tomo-e Gozen; 2: Kanata/HOWPol; 3: Kanata/HONIR; 4: DSO-14; 5: RRRT 6: TRT(SRO/SBO).
}
\end{deluxetable*}

\begin{deluxetable*}{cccccccc}
\movetabledown=3000mm
\tablenum{2}
\tablecaption{Swift/UVOT observations of SN~2020hvf\label{tab:Space Photometry Data}}
\tablewidth{0pt}
\tablehead{
\colhead{UT Date} & \colhead{Phase$^*$} & \colhead{$UVW2$} & \colhead{$UVM2$} & \colhead{$UVW1$} & \colhead{$U$} & \colhead{$B$} & \colhead{$V$}
}
\startdata
Apr 22.10 & -17.60 & 17.62 (13) & 17.58 (12) & 17.11 (14) & 16.58 (09) & 17.18 (09) & 16.74 (12) \\
Apr 23.17 & -16.54 & 17.00 (10) & 17.14 (15) & 16.13 (10) & 15.46 (06) & 16.27 (06) & 16.03 (08) \\
Apr 24.49 & -15.22 & 15.47 (06) & 15.24 (06) & 14.70 (06) & 14.37 (05) & 15.31 (05) & 15.26 (06) \\
Apr 25.69 & -14.04 & 14.62 (06) & 14.30 (06) & 13.88 (06) & 13.68 (04) & 14.71 (04) & 14.75 (05) \\
Apr 26.42 & -13.31 & 14.25 (05) & 13.91 (06) & 13.54 (05) & 13.38 (04) & 14.41 (04) & 14.40 (04) \\
Apr 29.07 & -10.67 & 13.65 (06) & -- & -- & 12.68 (04) & 13.61 (12) & 13.56 (04) \\
Apr 29.67 & -10.08 & -- & 13.20 (06) & 12.74 (06) & -- & -- & -- \\
May 05.52 & -4.26 & 13.60 (05) & 13.51 (06) & 12.53 (05) & 11.95 (04) & 12.81 (04) & 12.76 (04) \\
May 06.51 & -3.28 & 13.69 (06) & -- & 12.57 (05) & -- & -- & -- \\
May 07.71 & -2.09 & 13.78 (05) & 13.76 (06) & 12.69 (05) & 11.97 (05) & 12.70 (04) & 12.62 (03) \\
May 09.64 & -0.17 & 13.95 (07) & -- & -- & 12.16 (05) & -- & -- \\
May 11.14 & 1.33 & 14.11 (05) & 14.23 (06) & 13.02 (05) & 12.16 (05) & 12.68 (04) & 12.60 (03) \\
May 13.07 & 3.25 & 14.34 (06) & 14.46 (06) & 13.23 (05) & 12.41 (05) & 12.75 (04) & 12.65 (03) \\
May 21.24 & 11.37 & 15.28 (07) & 15.44 (07) & 14.23 (06) & 13.10 (06) & 13.20 (05) & 12.92 (04) \\
May 23.70 & 13.82 & 15.43 (07) & 15.61 (07) & 14.40 (06) & 13.30 (06) & 13.37 (04) & 13.06 (03) \\
Jun 01.53 & 22.59 & 16.23 (09) & 16.51 (10) & 15.32 (22) & 14.28 (08) & -- & 13.40 (04) \\
Jun 01.59 & 22.65 & 16.33 (09) & 16.42 (09) & 15.29 (08) & 14.32 (08) & 14.16 (08) & 13.41 (04) \\
Jun 04.18 & 25.23 & 16.45 (09) & 16.74 (10) & 15.64 (08) & 14.65 (08) & 14.41 (09) & 13.53 (04) \\
Jun 07.43 & 28.46 & 16.71 (11) & 16.77 (10) & 15.98 (10) & 15.03 (10) & 14.74 (06) & 13.81 (04) \\
Jun 10.22 & 31.24 & 16.86 (12) & 17.06 (12) & 16.12 (10) & 15.15 (10) & 14.92 (07) & 13.93 (05) \\
Jun 13.74 & 34.73 & 16.95 (11) & 17.13 (10) & 16.26 (09) & 15.44 (10) & 15.06 (04) & 14.03 (04) \\
Jun 16.80 & 37.77 & 17.15 (11) & 17.30 (11) & 16.32 (09) & 15.44 (10) & 15.12 (05) & 14.15 (04) \\
Jun 24.31 & 45.24 & 17.44 (11) & 17.58 (11) & 16.44 (09) & 15.57 (06) & 15.27 (05) & 14.42 (04) \\
Jul 01.08 & 51.97 & 17.57 (14) & 17.81 (14) & 16.80 (13) & 15.57 (07) & 15.37 (06) & 14.52 (05) \\
Jul 08.04 & 58.90 & 17.52 (13) & 17.92 (14) & 16.69 (11) & 15.65 (07) & 15.44 (05) & 14.69 (05) \\
Jul 15.08 & 65.90 & 17.68 (16) & 17.85 (16) & 16.92 (15) & 15.76 (09) & 15.51 (06) & 14.92 (07) \\
\hline
\enddata
\tablecomments{\\
The magnitudes are in the Vega system. Numbers in parentheses correspond to 1$\sigma$ statistical uncertainties in units of 1/100 mag.\\
$^*$ Days (rest frame) relative to the estimated date of the $B$-band maximum, 2020 May 9.8047.
}
\end{deluxetable*}

\clearpage

\end{document}